\documentclass[12pt,preprint]{aastex}

\newcommand{\simlt}{\lower.5ex\hbox{$\; \buildrel < \over \sim \;$}}

\slugcomment{draft date: \today}

\begin{document}

\title{The Dust cloud around the White Dwarf G 29-38. 2. 
Spectrum from 5-40 $\mu$m and 
mid-infrared photometric variability}

\author{William T. Reach\altaffilmark{1},
Carey Lisse\altaffilmark{2},
Ted von Hippel\altaffilmark{3,4},
Fergal Mullally\altaffilmark{5},
}

\altaffiltext{1}{Infrared Processing and Analysis Center, MS 220-6, 
California Institute of Technology,
Pasadena, CA 91125}

\altaffiltext{2}{Planetary Exploration Group, Space Department,
Johns Hopkins University, Applied Physics Laboratory, Laurel, MD}

\altaffiltext{3}{Physics Department, Siena College, 515 Loudon Road, Loudonville, NY 12211}

\altaffiltext{4}{Visiting Research Scientist, Department of Physics, Florida International
University, 11200 SW 8th St, Miami, FL 33199}

\altaffiltext{5}{Department of Astronomy, University of Texas, 1 University Station C1400, Austin, TX 78712}



\email{reach@ipac.caltech.edu}

\begin{abstract}
We model the mineralogy and distribution of dust around the white dwarf G29-39 using
the infrared spectrum from 1-35 $\mu$m.
The spectral model for G29-38 dust combines a wide range of materials based on spectral
studies of comets and debris disks.
In order of their contribution to the mid-infrared emission, the most abundant minerals 
around G29-38 are
amorphous carbon  ($\lambda<8$ $\mu$m),
amorphous and crystalline silicates (5--40 $\mu$m), 
water ice (10--15 and 23--35 $\mu$m), 
and metal sulfides (18--28 $\mu$m).
The amorphous C can be equivalently replaced by other materials (like metallic Fe)
with featureless infrared spectra.
The best-fitting crystalline silicate is Fe-rich pyroxene.
In order to absorb enough starlight to power the observed emission, the disk must either
be much thinner than the stellar radius (so that it can be heated from above and below) or
it must have an opening angle wider than $2^{\circ}$.
A `moderately optically thick' torus model fits well
if the dust extends inward to 50 times the white dwarf radius,
all grains hotter than 1100 K are vaporized,
the optical depth from the star through the disk is $\tau_{\parallel}=5$, and
the radial density profile $\propto r^{-2.7}$;
the total mass of this model disk is $2\times 10^{19}$ g.
A physically thin (less than the white dwarf radius) 
and optically thick disk can contribute to the near-infrared continuum only;
such a disk cannot explain the longer-wavelength continuum or strong emission features. 
The combination of a physically thin,
optically-thick inner disk and an outer, physically thick and moderately optically thin
cloud or disk produces a reasonably good fit to the spectrum and requires only silicates in the outer
cloud.
We discuss the mineralogical results in comparison to planetary materials.
The silicate composition contains minerals found from cometary spectra and
meteorites, but Fe-rich pyroxene is more abundant than enstatite 
(Mg-rich pyroxene) or forsterite (Mg-rich olivine) in G29-38 dust, in contrast
to what is found in most comet or meteorite mineralogies. 
Enstatite meteorites may be the most similar solar system materials to 
the G29-38 dust.
Finally, we suggest the surviving core of a `hot jupiter' as an alternative (neither cometary nor asteroidal) origin for the 
debris, though further theoretical work is needed to determine if this hypothesis is viable.
\end{abstract}


\section{Introduction}

The end state of stellar evolution for most types of star is a white dwarf.
Planetary material is common around main sequence stars \citep{backman,rhee,trilling}.
Part of the planetary system is expected to survive the violent, late stages of stellar
evolution \citep{debes02}.
Thus it is to be expected that planetary materials are present around white dwarfs.
But since white dwarfs are so under-luminous, this material is normally impossible to detect
via reflected light or thermal emission, if it is distributed like the Solar System (with
its inner portion destroyed out to at least the maximum radius of the red giant/AGB 
photosphere). Young, hot white dwarfs, where debris can be detected even at 100 AU from
the star \citep{suchu}, are the exception to this rule.
Planetary material can become detectable if it moves sufficiently close to a star, for example
via gravitational perturbations such as produce comets in the inner Solar System (from their
Kuiper Belt or Oort cloud homes) and
meteorites on the surface of Earth (from their asteroid belt homes).
\citet{jura03} explained how an asteroid can pass sufficiently close to a white dwarf to be tidally
disrupted, leaving a disk of dust near the star.
Sun-grazing comets in the solar system pass within similar distances.
The new and growing class of dusty, metal-rich white dwarfs 
(classified as DAZd, where D=white dwarf, A=mostly hydrogen atmosphere, 
Z=trace atmospheric metals, d=dusty) allow a unique window into planetary systems 
\citep{vonhippel}.

The first-known DAZd star, and the one with the brightest infrared excess, is 
Giclas 29-38 (WD 2326+049; G29-38 hereafter).
Its effective temperature has been estimated
in the range 11,800 to 12,100 K with surface
gravity $\log g$ in the range 8.14 to 7.90
\citep[e.g.][]{bergeron04,koester05}.
In \citet{holberg06}, the model corresponding to the
lowest estimated $T$ highest $g$ has radius 8100 km 
and mass .69 $M_\odot$, 
while the model corresponding to the highest $T$ and lowest $g$ 
has radius 9650 km and mass .55 $M_{\odot}$.
At wavelengths longer than 2.5 $\mu$m, the spectrum is completely dominated by 
non-photospheric thermal emission with color temperatures 300--900 K. 
In paper 1 \citep{paperone}, we measured the spectrum of this star from 5.5--14 $\mu$m,
revealing a remarkably strong emission feature from 9--11 $\mu$m that is typical
of small ($< 2 \mu$m radius) silicate grains. The brightness of the disk permits
spectroscopy at sufficient signal-to-noise to study the dust mineralogy. We have now
measured the infrared spectrum out to 35 $\mu$m, and in this paper we discuss dust
cloud models and compositions in more detail.

After presenting the new observations, analyzing the variability, and compiling the spectral energy distribution in
\S\ref{obssec}, we model the disk several different ways.
In \S\ref{mineralsec}, we use a wide mix of minerals, and a simple temperature distribution, to
determine likely constituents of the disk based on spectral features.
In \S\ref{modsec}, we present optically thin shell, moderately optically thick disk, and
optically thick thin-disk models.
In \S\ref{varsec}, we present constraints based on variability.
In \S\ref{discussion}, we tie together the evidence gleaned from comparing the models to
the observations.

\section{Observations\label{obssec}}

Table~\ref{obstab} shows a log of observations with the 
{\it Spitzer} Space Telescope
\citep{werner}. 
The Infrared Spectrograph (IRS) \citep{houck}
observations were performed in ``staring''  mode,
wherein the source was placed on each of two nod positions
on each of the slits and orders.
IRS data were processed with the S16 pipeline 
(with backgrounds subtracted by differencing nods)
and extracted
using optimal extraction in the {\it Spitzer} Science Center's data
analysis tool SPICE. The 4.5, 8, and 24 $\mu$m observations described in Paper 1
were supplemented by archival Infrared Array Camera (IRAC) \citep{fazio} 3.6, 4.5, 5.8, and 8 $\mu$m
observations.
Annular aperture photometry was performed on each IRAC basic calibrated image 
with a 4 pixel on-source radius and 8-20 pixel background annulus; array-location-dependent,
aperture-loss, and pixel-phase photometric corrections were applied; and the uncertainty
of each image's photometry was determined by combining uncertainties due to photon statistics and background removal \citep[see][]{reachcal}.

\subsection{Variability of the mid-infrared flux}

The variability of 
mid-infrared flux of G29-38 can be
constrained with the IRAC photometry.
\footnote{
We tested the IRS observations for variability, but there was no clear
evidence. The IRS observing strategy provided 6 samples along 60 sec ramps
in each subslit.
The flux varied by $< 4$\% among the shortest wavelength spectra, 
with no wavelength dependence detectable above the noise.}
Fluxes are shown in Table~\ref{iractab}.
The brightness of the star on each basic calibrated image
from each observing sequence was measured, corrected for
the array-location-dependent response, then a weighted
mean and statistical uncertainty in weighted mean computed.
The absolute calibration uncertainties are not included, 
because we are comparing fluxes from the same instrument;
IRAC photometry has been shown to be stable to 
better than 1\% \citep{reachcal}.
To obtain the highest signal-to-noise from
each image for repeatability,
we found an aperture radius of 4 pixels was optimal.
The signal-to-noise was 230, 2008, 90, and 130 at
3.6, 4.5, 5.8, and 8 $\mu$m.
At 4.5 \& 8 $\mu$m, the star was observed in two
independent observations just a few minutes apart,
yielding identical mean fluxes; the
flux differences are $-0.1\pm 0.2$\% and
$-0.1\pm 0.3$\%, respectively.
At 3.6 \& 5.8 $\mu$m, the star was observed with
different AORs separated by 13 months. 
The star was fainter in the second epoch, by
$-3.3\pm 0.4$\% at 3.6 $\mu$m
(and a statistically insignificant $-1.0\pm 0.8$\%
at 5.8 $\mu$m). The brightness difference at 3.6 $\mu$m
is likely due to short-term fluctuations in the star,
to which we turn now.

\begin{figure}[th]
\plotone{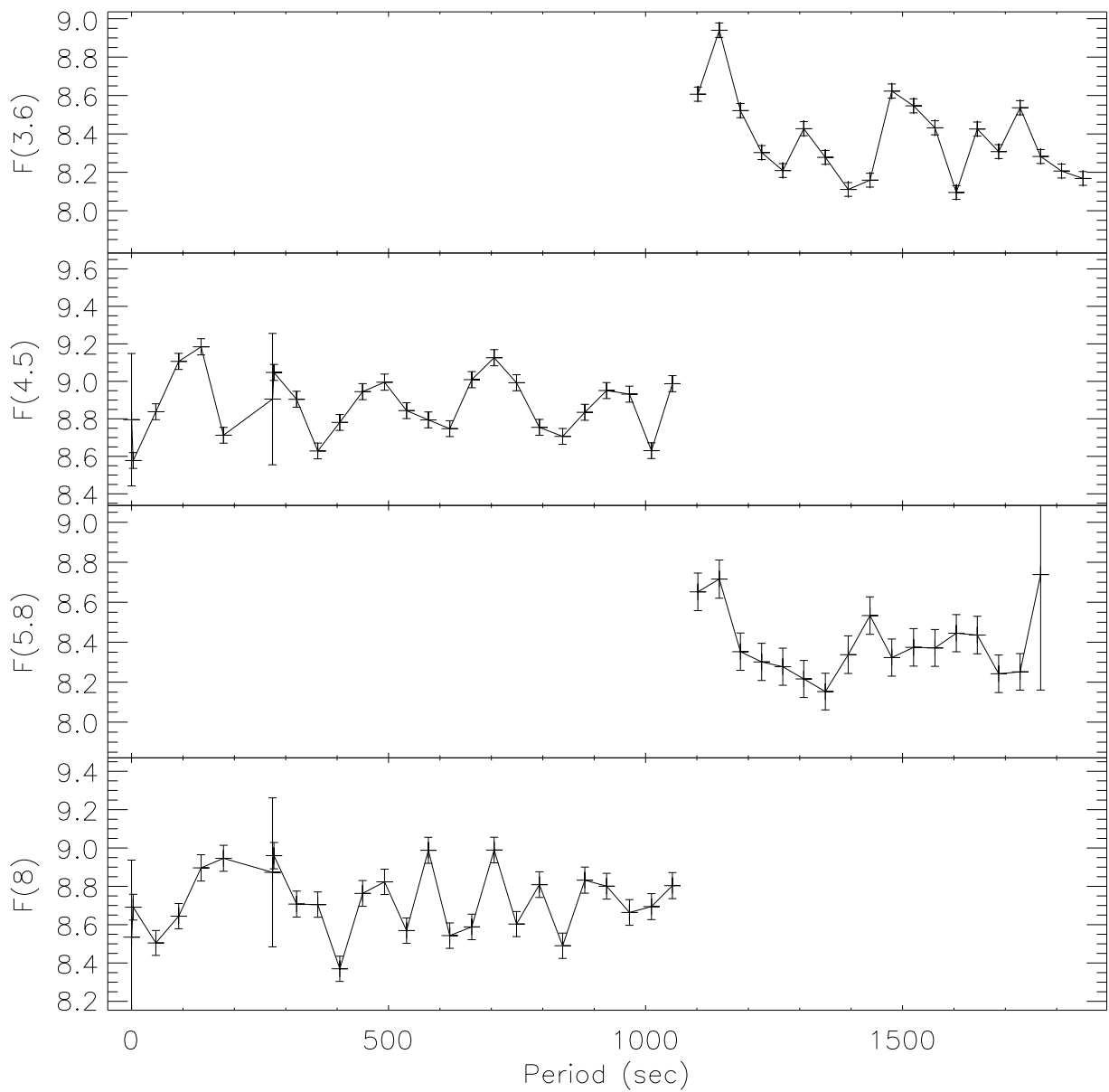}
\caption[sourcevar]{
Flux versus time for each of the IRAC channels
(from top to bottom: 3.6, 4.5, 5.8, and 8 $\mu$m wavelength).
The time begins with the first frame of the entire
sequence (which combines two AORs) at
2004 Nov 26 10:54:11.28 UT.
Each point is a flux measurement from a single image,
with the statistical uncertainty of the aperture
photometry performed as described by the
IRAC calibration procedure \citep{reachcal}.
Line segments simply connect the data points.
The high-error points at 4.5 and 8 $\mu$m are the
short frames taken just at the beginning of each
of the two AORs. All four plots are scaled to show the
same dynamic range (from -6\% to +8.4\%, centered
on the median).
\label{sourcevar}}
\end{figure}

The variability of the mid-infrared flux over shorter
timescales can be assessed from the individual images
taken in 2004 Nov. By the design of IRAC \citep{fazio},
the 3.6 and 5.8 $\mu$m channels observe simultaneously,
as do the 4.5 and 8 $\mu$m channels. The sequence of
events was as follows, with times given relative to the first
image at 2004-Nov-26 10:54:11.8 UT: 
\begin{itemize}
\item[00:00] the star is placed in the 4.5+8 $\mu$m field of view
and a short (1.2 sec) frame is taken,
\item[00:03] 5 consecutive dithers, consisting of a 30 sec frame
and a short telescope slew, are performed,
\item[04:34] the telescope is moved to re-center the star in the 
4.5+8 $\mu$m field of view and a short (1.2 sec) frame is taken,
\item[04:37] 19 consecutive dithers, consisting of a 30 sec frame
and a short telescope slew, are performed,
\item[18:22] the telescope is moved to center the star in the 
3.6+5.8 $\mu$m field of view,
\item[18:22] 19 consecutive dithers, consisting of a 30 sec frame
and a short telescope slew, are performed, then
\item[30:51] the sequence is complete.
\end{itemize}
Figure~\ref{sourcevar} shows the flux versus time.

\begin{figure}[th]
\plotone{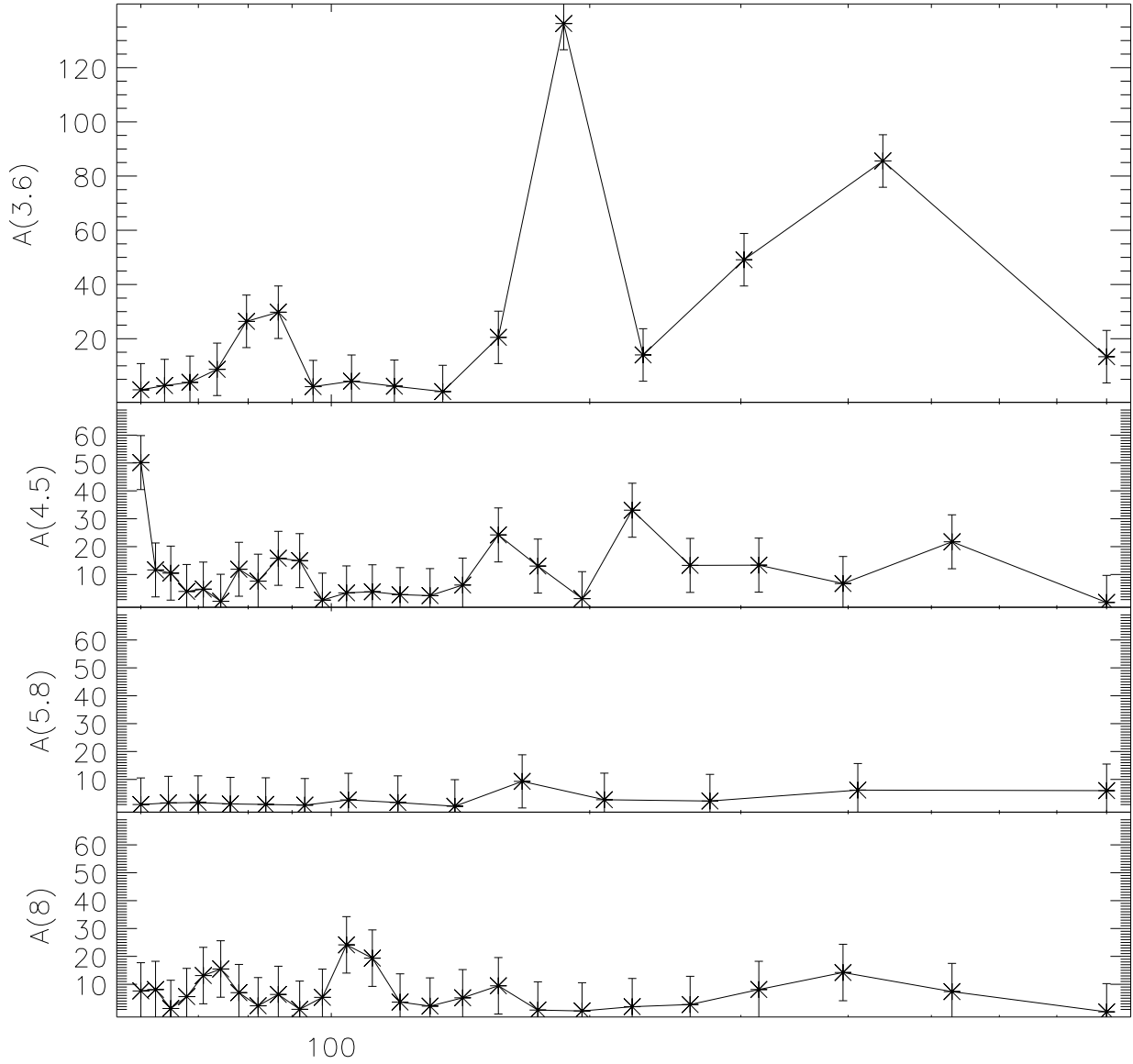}
\caption[sourceper]{
Scargle periodograms of the time series from the IRAC
2004 Nov 26 observations. Each panel has the same scale
both horizontally and vertically; the 3.6 $\mu$m panel
is twice as large because the amplitudes are much higher.
The periods ($2\pi/\omega$, where $\omega$ is the angular
frequency) are displayed on a logarithmic stretch.
\label{sourceper}}
\end{figure}

The scatter in the photometric measurements is significantly
larger than their uncertainties, especially at
3.6 $\mu$m. To assess the harmonic content of the time
series, we computed the periodogram as described
by \citet{scargle}. Figure~\ref{sourceper} shows the
periodograms at the 4 IRAC wavelengths.
The time series cannot be described by a simple period, 
but instead contain variation on a 
range of timescales.
This is characteristic of ZZ Ceti variable stars
\citep{kleinman}.
The peak in the 3.6 $\mu$m amplitude at 190 sec is
highly significant ($14\sigma$), as is the
harmonic content at 300-440 sec. None of the other
harmonic content is significant, at the sampling
frequency and signal-to-noise level of these
observations, except some power ($3.3\sigma$) at $\sim 220$ sec
at 4.5 $\mu$m.

The flux variations seen with IRAC are similar to those seen
at Palomar by \citet{graham}, who observed
simultaneously at wavelengths dominated by the photosphere
(B, J) and the infrared excess (K-band).
They found fluctuations at periods
181 and 243 sec in K-band, with no corresponding 
ones at J-band. Our IRAC 3.6 $\mu$m results confirm 
the significant fluctuations at $\sim 190$ sec.
The amplitude of the fluctuations $\sim 4$\%
at 3.6 $\mu$m, while \citet{graham} found
2.5\% variations at 2.2 $\mu$m.
The stellar photosphere must be subtracted before
interpreting these results.
At 2.2 (3.6) $\mu$m, the photosphere contributes
66\% (18\%) of the total flux. If we assume
the infrared photosphere is constant, and the
fluctuations are due to the disk, then the
amplitude of disk fluctuations 
at 2.2 (3.6) $\mu$m is 7\% (5\%),
i.e. very similar, with K-band possibly slightly
higher in amplitude.
Fluctuations at 4.5, 5.8, 8 $\mu$m are not detected
in the periodogram,
with upper limit $\sim$ 5\%, 5\%, 3\% of the total
flux, or 5\%, 5\%, 3\% of the infrared excess
above photosphere ($2\sigma$ limits). The upper limit
at 8 $\mu$m, and the trend such that fluctuations are
most significant at 2.2 and 3.6 $\mu$m, with a rapid 
decrease in amplitude at longer wavelengths, are significant
and indicate that the fluctuations are due to an
emitting region with a relatively high 
($> 1000$ K) color temperature.

\def\oldnote{
{\bf NOTE:} discuss variability WITHIN the IRS observations either here or above...if full 8\% 
observed variability applies across the wavelength range
 then within
the spectra there could have been variations; certainly between
subslits...
original SL1+2 observations were 3 cycles of 60 sec each subslit so they took ~400 sec per subslit to perform; this is long enough that the IRAC-observed variations would have occurred
}

\subsection{Compilation of the spectral energy distribution}

In order to model the mineralogy and dust distribution around G29-38, we
need to combine the observed data from the near- through mid-infrared,
correcting for calibration errors as well as source variability. This is
in fact not possible, since the temporal sampling of the photometry is 
inadequate, and the observations are not contemporaneous. We therefore
must assemble the various portions of the spectrum in such a way as to make
them most plausibly `connected'. In principle, there is a scale factor for
each wavelength range that depends on the epoch. The amplitude of the scale
factor, due to source variability,
may reach up to $\sim 10$\% in the near-infrared and should be less
than 5\% at longer wavelengths. Calibration uncertainties range from $\sim 3$\%
in the optical to 5\% for IRAC and MIPS to 10\% for IRS. 
The range of plausible scale factors is then 5-10\% across all wavelengths.

We first ensured that the spectroscopy and photometry were in accord.
The IRAC fluxes must be corrected for the spectral shape of the source first.
For the IRAC 8 $\mu$m channel, which is fully covered by the IRS spectrum,
we integrated the IRS spectrum appropriately 
\citep{reachcal} over the passband to calculate
the color-correction $K=1.16$ 
(where the corrected flux is the `observed' flux from the IRAC calibration
divided by $K$)
and applied this correction to the photometry.
The IRAC 8$\mu$m flux density can then be compared directly to the IRS spectrum:
the flux at the nominal wavelength 7.872 $\mu$m is 7.19 mJy from IRAC and 7.30 mJy
from IRS, a deviation of only 1.5\%, well within the calibration error budget.
The IRAC 3.6, 4.5, and 5.8 $\mu$m flux density required only very small color corrections:
the spectrum has a color temperature $\sim 1100$ K at these wavelengths, 
for which the corrections are 0.995, 0.996, and 0.996, respectively.
The absolute fluxes from IRAC and IRS at 5.831 $\mu$m are 7.76 and 7.54 mJy, showing
the instrument cross-calibration is good.
The observed MIPS 24 $\mu$m flux was divided by a color correction of 0.97 to
account for the wide spectral response of the filter and detector.
For IRAC and IRS, we therefore find that there is no need for a relative re-scaling;
that is, the measured flux densities averaged over their exposure times at their
epoch of observation agree well, whether due to averaging of {\it or} lack of
variability.

To remove the white dwarf photosphere from the spectra, we use
a model atmosphere for $T=12,000$ K
and $\log g=8$ covering 0.35--60 $\mu$m (courtesy D. Koester),
normalized using
2MASS photometry ($J=13.132\pm 0.029$, $H=13.075\pm 0.022$, 
$K_s=12.689\pm 0.029$) and optical spectrophotometry 
from Palomar \citep{greenstein}.
In fact, the optical and 2MASS photometry cannot both agree with the model
spectrum. If we normalize the white dwarf model at the 2MASS J band, then
the model would under-predict the visible flux by 8\%.
The near-infrared spectrum obtained by \citet{kilicspec} with the IRTF also disagrees with
the 2MASS photometry; if we normalize the white dwarf model by the 2MASS J-band
flux, then we under-predict the IRTF spectrum by 18\%.
The low flux observed by 2MASS is likely due to variability of the white dwarf.
The shape of the IRTF spectrum follows the model closely at wavelengths
shorter than 1.6 $\mu$m; the infrared excess is evident at wavelengths beyond 1.7 $\mu$m.
To obtain a joined spectral energy distribution we proceed as follows:
(1) normalize the white dwarf model to match the optical photometry,
(2) scale the 2MASS photometry by a factor 1.087 to match the model at J band,
(3) rescale the IRTF spectrum by 0.92 to match the white dwarf model at 0.8--1.3 $\mu$m,
(4) rescale the 2MASS photometry and IRTF spectra by a factor of 1.2 so that the red end of
the IRTF spectrum (2.4 $\mu$m) is plausibly consistent with the IRAC 3.6 $\mu$m
photometry.  
(Plausible consistency here was defined as allowing a blackbody to pass through the
IRAC 3.6 $\mu$m and 2MASS K$_{s}$ photometry for color temperatures 800--1200 to
within the error bars, and keeping the IRTF and 2MASS photometry consistent at 2.17 $\mu$m.)
The uncertainties were taken to be a root-sum-square combination of the measurement
uncertainties and 10\% of the brightness of the photosphere, to account for
scaling uncertainties in the photospheric subtraction.

\begin{figure}[th]
\plotone{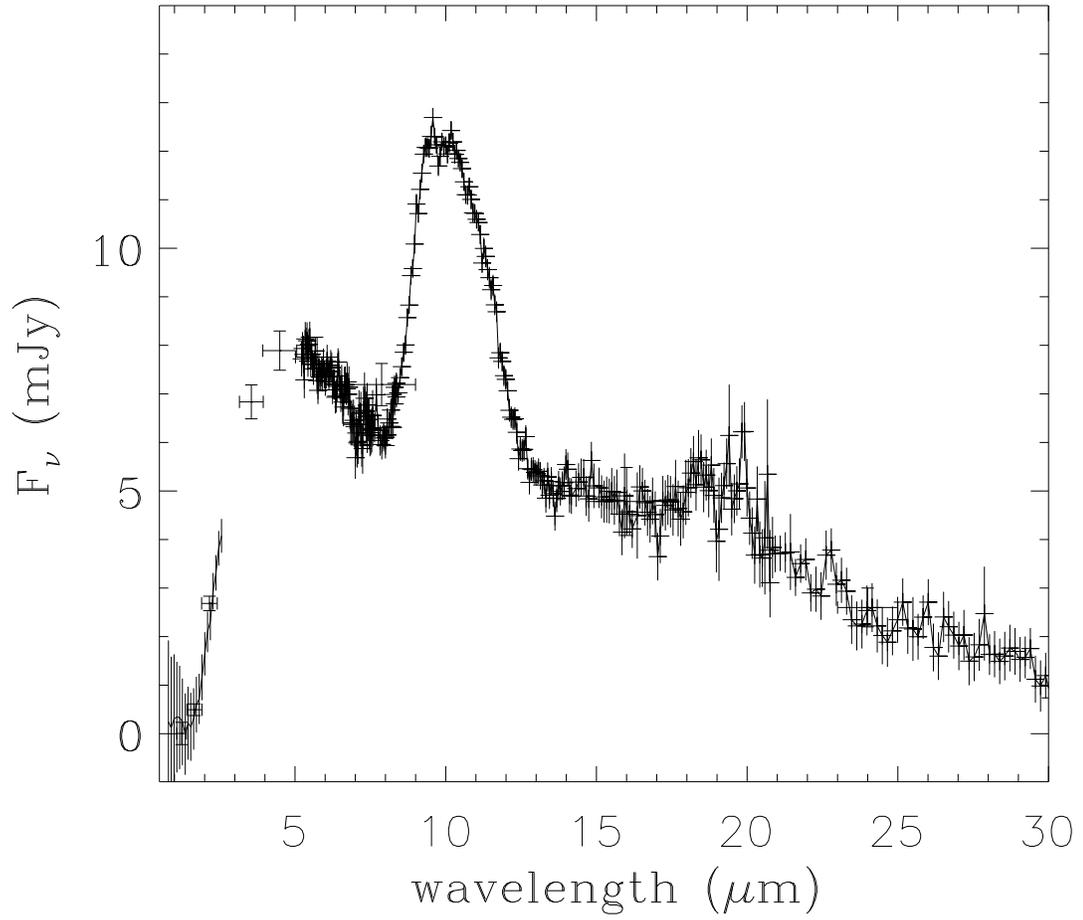}
\caption[ffit]{
Combined spectra and photometry
(using 2MASS, IRTF, IRAC, MIPS, and IRS)
of G29-38 with stellar photosphere removed.
\label{ffit}}
\end{figure}

Figure~\ref{ffit} shows the combined spectrum of the infrared emission after
subtracting the stellar photospheric emission.
The combined scalings yield near-infrared photometry and spectrum
effectively at the same epoch (in terms of variability) 
as the mid-infrared photometry.

\section{Mineralogy\label{mineralsec}}

The composition of the dust around G29-38 was determined by comparing the spectra to those of
a set of constituent materials, using a fitting method that has
been used for comet Hale-Bopp, the Deep Impact ejecta from
comet 9P/Tempel 1, and the dust around the stars HD 100546, HD 69380, and HD 113766
\citep{lisse06,lisse07a,lisse07b,lisse08}. Details of this model are given in
\citet{lisse08}.
In this section, we model the emitting region as an optically thin dust torus.
We apply this simple model to G29-38 to allow direct comparison to dust in other astronomical
systems, but we will revisit the mineralogy below when testing more sophisticated
models.

\subsection{Particles sizes and temperatures}
A toroidal model is motivated by the single temperature distributions found for many of the stars studied by 
\citet{beichman06a} and \citet{chen06} as well as the narrow dust structures found in many of the HST images of debris disks 
\citep{kalas05,kalas06}. The best-fit temperature of the smallest dust particles (0.1--1 $\mu$m), which superheat substantially above Local Thermal Equilibrium (LTE), is $T_{max} = 890$ K for the olivines, 850 - 890 K for the pyroxenes, and 930 K for the amorphous carbon. The largest particles in our calculation, with radius 1000 $\mu$m, are set to $T_{LTE}$ = 600 K, and dust of intermediate size is scaled between the two extremes \citep{lisse06}. Using the Tempel 1 ejecta temperatures as a guide (where we found olivines at 340 K and amorphous carbon at 390 K at 1.51 AU from the Sun, where $T_{LTE}$ = 230 K), and allowing for an G29-38 stellar luminosity that is 
$2\times 10^{-3} L_{\odot}$, we estimate a location for the hot dust $\sim 150$ stellar radii
from the white dwarf. The location of colder dust, capable of supporting a stable water ice component at 200 K, is at $> 10^3$ stellar radii ($>8\times 10^{11}$ cm). 
While differing by at least a factor of 9, this range of distances is still small, suggesting material in tight orbit around the WD.
The best-fit single continuum temperature to the 
7--35 $\mu$m Spitzer spectrum is 950 K, close to the amorphous carbon temperature. As seen in the Tempel 1 ejecta from the Deep Impact experiment \citep{lisse06}, the amorphous carbon dominates the continuum behavior as it has a featureless emissivity and is the hottest material, thus contributing most to the short-wavelength emission. 

The best-fit size distribution
$dn/da\propto a^{-3.7\pm 0.2}$ argues for dust surface area dominated by small particles, but
dust mass dominated by large particles.
A system in collisional equilibrium would demonstrate a PSD  $\propto a^{-3.5}$ \citep{dohnanyi,durda97};
for ``real'' systems a size distribution even steeper than $a^{-3.5}$ at small sizes is expected in a collisional cascade 
because of the dependence of particle strength on size  \citep{obrien03}.

The total dust mass required to explain the Spitzer IRS spectrum 
(i.e., mass in particles of 0.1--10 $\mu$m in size that contribute appreciably to the $\chi^2_\nu$ value of the fit to the infrared spectrum)
is $\sim 2 \times 10^{19}$ g.
Extrapolating (using the best-fit size distribution)
from a maximum size of 10 $\mu$m to 1 cm would
increase the mass by a factor of 8.
A total cloud mass of order $10^{19}$ g compressed into a solid
sphere of average density 2.5 g~cm$^{-3}$ (i.e. rocky silicate material)
would have a radius 
of 10 km, equivalent to a single, small asteroid or large comet.
The surface area of {\it detected} particles is 
$5 \times 10^{22}$ cm$^{2}$. 
If this dust is in an annulus of inner radius $7\times 10^{10}$ cm 
($90 R_{*}$) extending to twice that radius, 
then the areal filling factor of grains viewed from above the disk
is of order unity. 
The optically-thin model is thus unlikely to apply, although most of the observed emission
(in particular the spectral features) must arise from the optically thin parts of the disk.
We address the optical depth effects in the modeling sections below, but we proceed first
(with
caution) to discuss the mineralogy from the optically thin fits.

\subsection{Dust Composition}

Over 80 different species were tested for their presence in the spectra. 
The material spectra were selected by their reported presence in meteorites, in situ comet measurements, YSOs, and debris disks \citep{lisse06}. 
Consultations with members of the {\it Stardust} team, and examination of the interplanetary dust particle (IDP) literature \citep[cf. review by][]{bradley02}, and the astrominerological literature \citep[cf. review by][]{molsterwaters} pointed to the most likely mineralogical candidates to be found in the dust. Materials with emissions matching the strong features in the Spitzer emissivity spectra were also tested.
The list of materials tested against the {\it Spitzer} spectra included multiple silicates in the olivine and pyroxene class (forsterite, fayalite, clino- and orth-enstatite, augite, anorthite, bronzite, diopside, and ferrosilite); phyllosilicates (saponite, serpentine, smectite, montmorillonite, and chlorite); sulfates (gypsum, ferrosulfate, and magnesium sulfate); oxides (various aluminas, spinels, hibonite, magnetite, and hematite); Mg/Fe sulfides (pyrrohtite, troilite, pyrite, and niningerite); carbonate minerals (calcite, aragonite, dolomite, magnesite, and siderite); water ice, clean and with carbon dioxide, carbon monoxide, methane, and ammonia clathrates; carbon dioxide ice; graphitic and amorphous carbon; and 
polycyclic aromatic hydrocarbons (PAH).
Of these materials, a small, unique subset was found necessary to properly fit the Spitzer data. 
Sources for the data included: for {\it silicates},
 the Jena Database of Optical Constants for Cosmic Dust\footnote{http://www.astro.uni-jena.de/Laboratory/OCDB} \citep{dorschner,jaeger98,jaeger03}, 
the Mars Global Surveyor Thermal Emission Spectrometer database\footnote{http://tes.asu.edu}, and W. Glaccum (private communication), as well as emission spectra 
from \citet{koike} and \citet{chihara};
for {\it carbonates}, \citet{kemper02};
for {\it sulfides}, \citet{keller} and \citet{kimura};
for {\it amorphous carbon}, \citet{edoh};
for {\it PAH}, \citet{lidraine}. 

We determined the reduced $\chi^{2}$ for the model fit to the data for 
thousands of combinations of minerals. All models with reduced chi-squared  values large than the 95\% confidence limit were excluded from consideration. 
The range of constituents abundances was determined by varying the amount of a material, and finding where the model exceeded the 95\% confidence limit. 
Only constituents with  abundances significantly above zero were included in the final best fit. 
Table~\ref{minfittab} shows the composition of the best-fit model. For each entry, 
we also show the reduced $\chi^{2}$ for the model if that constituent is deleted.
Upper limits for non-detected species are included in Table~\ref{minfittab} based on
the 95\% confidence level amplitude of the constituent when it is included in the fit.

\begin{figure}[th]
\plotone{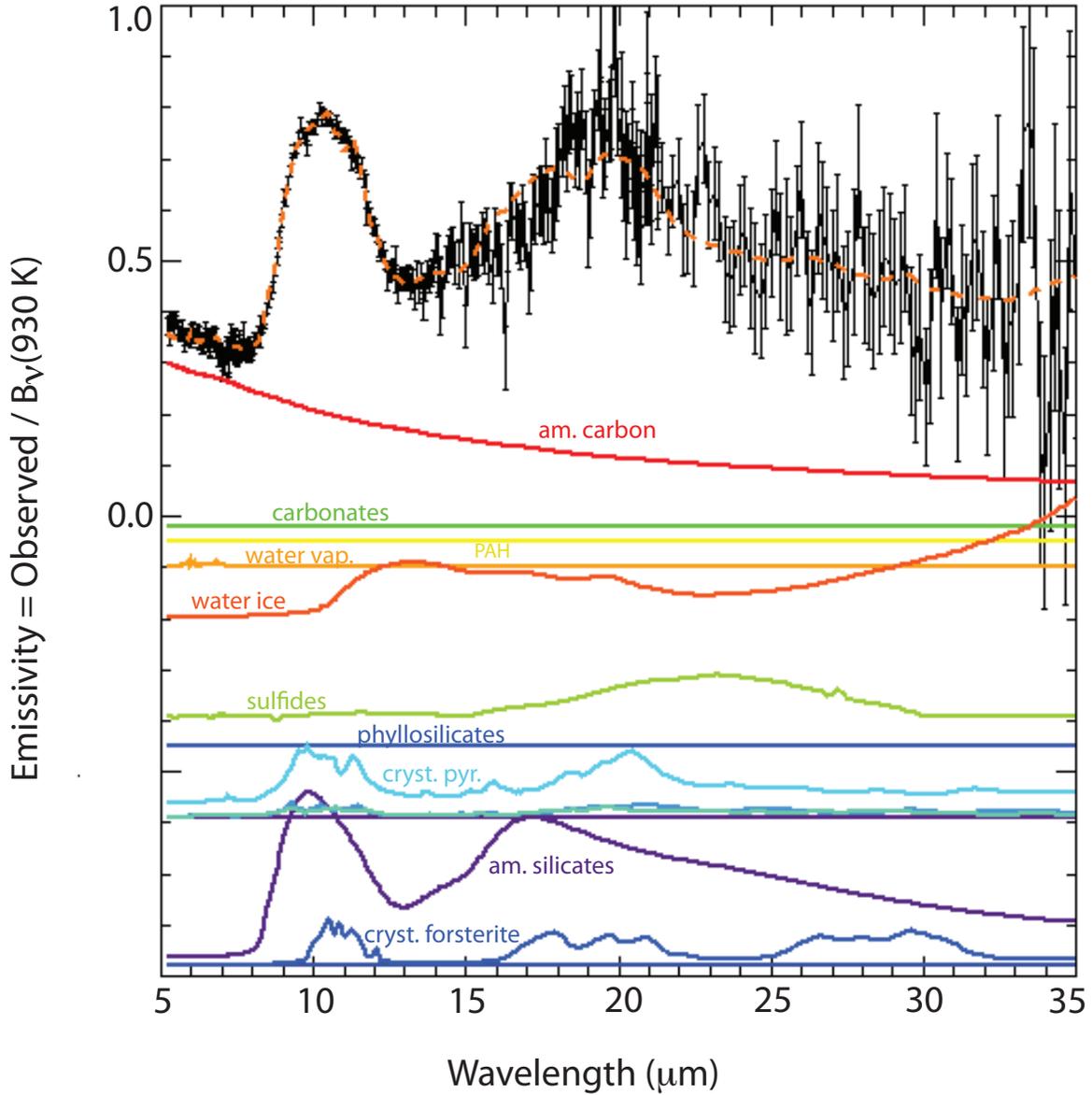}
\caption[specmod]{
Emissivity spectrum of the dust around G29-38. The observed
spectrum has been divided by a 930 K blackbody and fitted with
a linear combination of 12 minerals. The shape of each mineral's
emissivity, normalized by its fitted amplitude to the
G29-38 emissivity, is shown separately by a colored line
(offset vertically for clarity):
red = amorphous carbon,
bright green = carbonates (zero amplitude),
yellow = PAH (zero amplitude),
light orange = water vapor (zero amplitude),
deep orange = water ice,
olive green = sulfides, represented here by niningerite,
blue = phyllosilicates (zero amplitude),
light blue = crystalline pyroxenes (ferrosilite,
diopside, and orthoenstatite, in order of 20 $\mu$m amplitude),
purple = amorphous olivine, and
dark blue = crystalline olivines (forsterite and fayalite, in order of 20 $\mu$m amplitude).
\label{specmod}}
\end{figure}

\begin{figure}[th]
\plotone{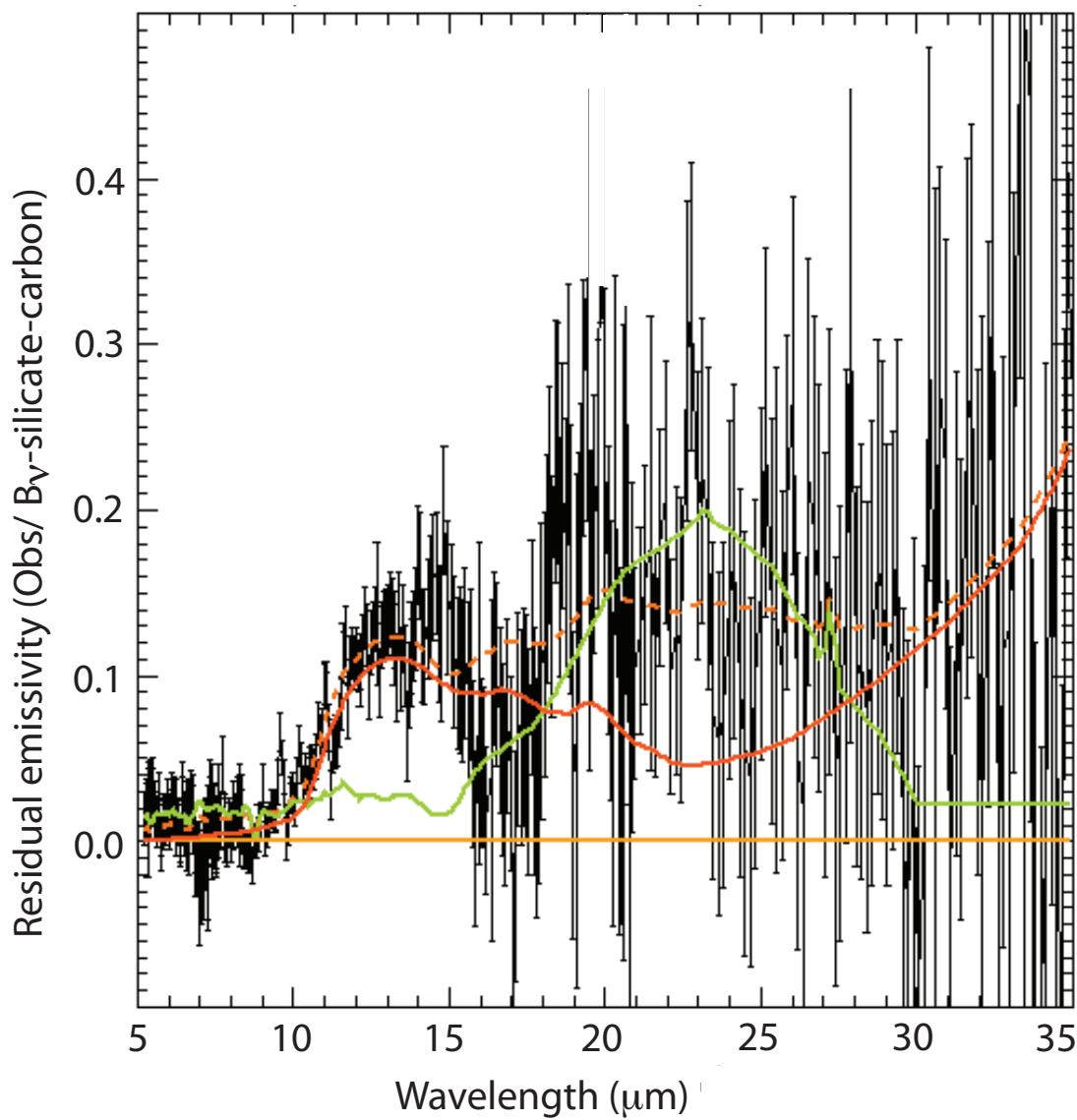}
\caption[specresid]{
Emissivity spectrum as in Fig.~\ref{specmod}, but after removal of
the best-fitting silicates and carbon.  
The residuals are well fit by a combination of water ice (deep orange, dashed and solid are two 
temperatures) and metal sulfides (olive green). 
The water ice is at temperature $~200$ K, and cannot be in the same location as the 930 K dust.
Whereas the dust is $\sim 10^{11}$ cm from the star, the water ice must be 
further $\sim 10^{13}$ cm, at the outer edge of an extended disk.
\label{specresid}}
\end{figure}

Figure~\ref{specmod} shows the best-fit spectral model for G29-38. 
In order of their contribution to
the mid-infrared emission, the contributing minerals are
amorphous carbon;
amorphous olivine;
crystalline silicates 
ferrosilite (FeSiO$_{3}$),
fayalite (Fe$_{2}$SiO$_{4}$), 
diopside (MgCaSi$_2$O$_6$), and
enstatite (MgSiO$_{3}$);
and metal sulfides (Mg$_{10}$Fe$_{90}$S); and
water ice.
Figure~\ref{specresid} shows the spectral model after removal
of the best-fitting amorphous silicates that dominate the overall
spectral shape. The fit to the IRS data (5.2--35 $\mu$m) is excellent,
with $\chi^{2}_{\nu}=1.03$.

The composition of the dust surrounding G29-38, as determined by modeling of the Spitzer IRS
spectrum, is unusual when compared to circumstellar dust in other environments. 
Amorphous olivine is present, as in the other systems that have been modeled with the same technique. 
But the relative lack of crystalline olivine, and
the strong prevalence of Fe-rich silicates, make the G29-38 material
distinct from other systems.
The mix of crystalline pyroxenes and amorphous
olivines may be indicative of `aged' dusty material. 
No PAHs, carbonates, or water gas are
seen, reflecting a total lack of primitive nebular
material. Some metal sulfides appear to be present, suggesting temperatures as low as $\sim$600
K in the observed dust; 600--700 K is the temperature range for vaporization/condensation of
ferrosulfides. Given that the best-fit continuum temperature for the spectrum is ~930K
(dominated by the short wavelength emission from 0.1 - 1.0 $\mu$m amorphous carbon particles),
either the carbon is superheated beyond the temperature of the metal sulfides, which may be
present in larger, cooler particles, or there is a distribution of dust locations and effective
temperatures, i.e. a disk-like structure. 

\subsection{Water ice} 
The detection of water ice emission in the spectrum is curious. 
In a vacuum, water ice sublimates at temperatures above 
$\sim$200 K. 
Thus water ice cannot be in direct physical contact with the rest of the hot dust detected 
in the infrared spectrum.
Either the dust must be continuously created, or it must reside in a physical location removed from the rest of the dust reservoir. 
Since no water vapor is detected at 6 $\mu$m, where there are strong features
that should be detected if water were present in significant quantities 
\citep{lisse06,woodward07}, 
no appreciable ongoing sublimation can be occurring. 
Water vapor is ionized within $\sim 10^{3}$ sec,
if we scale from the lifetime of water at 1 AU from the Sun
\citep[$\sim 10^6$ s;][]{schleicherahearn84} according to the luminosity of G29-38 ($2\times10^{-3} L_{\odot}$).
There would still be a steady state amount of water vapor present
if the ice is sublimating.
Thus it appears that the water ice is at a large remove from the rest of the warm circumstellar dust, and the water ice is found in the IRS beam (i.e. within 
$5\times 10^{14}$ cm [$7\times 10^{5} R_{*}$] of the star).

In paper 1, we showed that a range of temperatures is required to match the
photometry out to 24 $\mu$m, with a two-temperature fit having 890 and 290 K color
temperatures. Thus we already suspected either a continuous range of dust temperatures or
the presence of two dust reservoirs, one hot and one cold.
For G29-38, the ice (or ice-coated grains) must be located at 
$>10^{12}$ cm ($2000 R_{*}$)from the star, which,
while farther than the dust that dominates the mid-infrared spectrum, is still very close to the
star. 
Cold, icy dust around other stars, at locations comparable to the Solar System's Kuiper Belt,
is common, being present around 15--20\% of all stars \citep{bryden06}.
The ice detected around G29-38 is  closer to the star than the far-infrared debris disks at 
$\sim 100$ AU ($10^{15}$ cm) radius
commonly seen around main sequence stars \citep{bryden06}, and the present location
of the icy grains would have been within the star
when G29-38 was in its red giant or AGB stage.
The source of this cold, presumably icy, material is likely to be the same
parent bodies that create the dust closer to the star. We will return to the 
possible parent body natures in the Discussion section, but we should point out
now that presence of H$_{2}$O is well established in asteroids (especially outer belt),
meteorites (as water of hydration as well as signatures of aqueously altered mineralogy),
comets, and planets. The H$_{2}$O in any of these bodies would freeze if it were
liberated by disintegration of these bodies and survive out to $>10^{12}$ cm from the 
white dwarf, owing to its small luminosity.

\def\extra{
THIS MATERIAL LOOKS IRRELEVANT TO G29_-38!!!

Extrapolating the best-fit model spectrum longwards of 35 $\mu$m 
using an assumed absorption efficiency 
$Q_{abs} = Q_{abs}(35 \mu{\rm m})/\lambda$, 
where $\lambda$ is the wavelength, as a lower limit and 
$Q_{abs}=$constant as an upper limit, we estimate a 
70 $\mu$m flux of 5 ± 4 mJy, while the Spitzer MIPS measurements of Beichman et al. (2005) yield  a value of 7 ± 3 mJy for the 70 um excess. We thus concur with conclusion that there is no large amount of cold (i.e., Kuiper Belt dust) contributing to the excess emission detected by Spitzer, and that the emission is dominated by the single debris disk at T~ 400 K .
}

\section{Distribution of material around the star\label{modsec}}

\begin{figure}[th]
\plotone{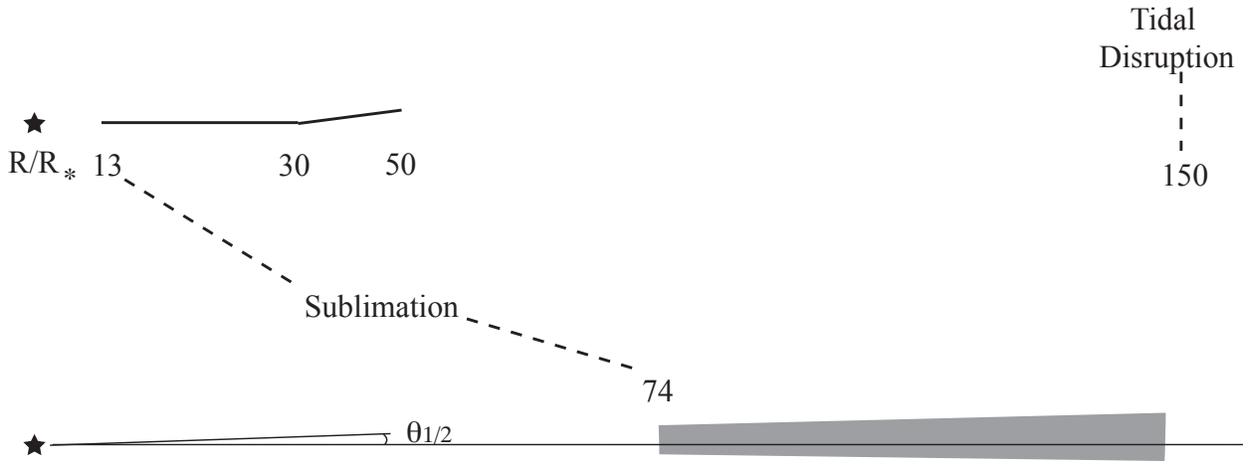}
\caption[g29v3]{
Cartoons illustrating possible geometries for the G29-38 disk.
The physically thin plus warp model (top) is based on \citet{jura07}'s model for the similar white dwarf GD 362.
The physically thick (bottom) is based on \citet{paperone}
and elaborated in \S\ref{modsec}. 
Distances from the star are labeled, in units of stellar radii.
The distances from the star for
grain vaporization (greybody grains at 1200 K)
and tidal disruption (Roche limit for a
solid body to be tidally disrupted by the white dwarf with
radius $7.5\times 10^{8}$ cm and mass 0.69 $M_{\odot}$)
are indicated.
\label{cartoon}}
\end{figure}

Two possible disk models are considered; these are
illustrated in Figure~\ref{cartoon} and discussed in
turn in the following sections. The models will be fitted
to a combination of the {\it Spitzer} spectroscopy and photometry, 
the 2MASS photometry, and the near-infrared
spectrum, after removal of the photosphere model
and scaling to a common epoch (\S2.2).

\subsection{Physically thick disk}

If the optical depth through the dust cloud around G29-38 is optically
thin at the wavelengths of the observed emission, then the 
spectrum is determined by simply integrating the density distribution,
weighted by the Planck function at the local temperature,
through the cloud. If the cloud is optically thin at
visible wavelengths, where the spectrum of the white dwarf peaks, 
then the dust heating is simply determined by the integrated
absorption of the star's distance-diluted spectrum by each grain.
We can place some constraints on whether such a model can apply. 
First, the cloud can only emit as much energy as it absorbs from the star.
Let us consider a flattened torus, defined in
spherical coordinates $r, \theta$ (with $r$ distance from the 
star, $\theta$ the angular separation from the equatorial plane)
as having non-zero 
density for $R_{1}<r<R_{2}$ and $\theta<\theta_{\frac{1}{2}}$. 
Since the dust luminosity is $f=3$\% of the star luminosity, the
opening half-angle of the torus must be at least 
$\theta_{\frac{1}{2}}>f/2$ radians, i.e. $\theta_{\frac{1}{2}}>0.8^{\circ}$. A thinner torus 
simply cannot intercept enough starlight to emit the observed
luminosity. (The constraint does not apply to a disk thinner than the
radius of the star, nor to a warped disk, as discussed in following sections.)

Of critical importance to the radiative transfer for calculating the emergent
spectrum from the disk is its optical depth in the infrared.
For a torus with radial mass density profile 
$\rho = \rho_{1} (r/R_{1})^{-\alpha}$,
the total cloud mass 
\begin{equation}
M = 4\pi \theta_{\frac{1}{2}} \rho_{1} R_{1}^{3} f_{\alpha},
\end{equation}
where
\begin{eqnarray}
f_\alpha &=& \frac{(R_2/R_1)^{3-\alpha}-1}{3-\alpha}  \,\,\,\,\,\,\,\,({\rm for}\,\alpha\neq3)\\
&=& \ln(R_{2}/R_{1})  \,\,\,\,\,\,\,\,\,\,\,\,\,\,\,\,\,\,\,({\rm for}\,\alpha=3).
\end{eqnarray}
The vertical optical depth at the inner radius is
\begin{equation}
\tau_{\perp1} = \frac{3 Q_{IR} M}{16\pi \rho_{d} a R_{1}^{2} f_\alpha}
\end{equation}
where $a$ is the particle radius and $Q_{IR}$ is the the absorption
efficiency averaged over the thermal emission spectrum.
Using the cloud mass radius from paper 1 ($R_{1}=1 R_{\odot}$) and
scaling the mass in units of $10^{18}$~g, 
$\tau_{\perp1} = 0.03 M_{18} Q_{IR} (a/\mu{\rm m})^{-1}$.
For particles smaller than the wavelength, the absorption efficiency
can be approximated as $Q_{IR}\simeq 2\pi a/\lambda$, 
so the optical depth for thermal emission at wavelength $\lambda$
is $\tau_{\perp1,\lambda}=0.2 M_{18} (\lambda/\mu{\rm m})^{-1}$,
independent of particle size.
For thermal emission at 10 $\mu$m, one can neglect radiative transfer out of the disk
only if $M<5\times 10^{19}$ g. 

Dust heating is determined by the propagation of starlight
through the disk. The optical depth 
from the disk inner boundary to a distance twice as far from the star is
\begin{equation}
\tau_{\parallel1}= \frac{\tau_{\perp1} Q_{opt}}{\theta_{\frac{1}{2}} Q_{IR}} \frac{1-2^{-\alpha}}{\alpha-1}.
\end{equation}
For particles larger than 0.1 $\mu$m, $Q_{V}\simeq 1$ for absorption of
starlight, and the optical depth from the star to the disk interior is
$\tau_{\parallel1}=2.5 M_{18} (\theta_{\frac{1}{2}}/0.8^{\circ})^{-1} a_{\mu{\rm m}}^{-1} (R_1/R_\odot)^{-2}$.
Using the mass and radius from paper 1, we find the cloud is optically
thin for starlight propagation (and for
thermal emission) as long as $\theta_{\frac{1}{2}}a_{\mu{\rm m}}> 2^{\circ} M_{18}$.
The presence of the silicate emission feature further requires the particle size $<2 \mu$m, so
for a disk that is optically thin to propagation to starlight, we
require $M_{18} < \theta_{\frac{1}{2}}$.
The cloud must be at least as massive as derived in paper 1 (where light
was allowed to propagate through the disk unimpeded), so we finally obtain the constraint
$1 < M_{18} < \theta_{\frac{1}{2}}$ for an optically thin disk.
If the disk is more massive than the upper limit, it becomes
optically thick to propagation of starlight, the temperature 
decreases more rapidly with distance from the star, 
and the cloud is more dominated by its inner edge. 
We will refer to this case as 
moderately optically thick, with $0.5<\tau_{\parallel}<10$ so
the cloud is optically thin to vertical propagation of 
the observed infrared emission but
optically thick to radial propagation of starlight (\S\ref{modthicksec}).

\def\extra{
The particle size can be constrained somewhat, without resorting to
detailed models. In order to produce the 
9--11 $\mu$m silicate
emission feature, we require $Q[10\,\mu{\rm m}]<1$, 
with the absorption efficiency $Q\simeq 2\pi a/\lambda$ for small particles, so
$a_{\mu{\rm m}} < 2$. Larger particles will certainly be present,
but the ones producing the mid-infrared emission feature must satisfy 
this criterion. 
Particles smaller than 0.1 $\mu$m will both absorb and emit inefficiently. The temperatures of such particles will be boosted over
those of greybody grains by a factor 
$T/T_{bb}\simeq (\lambda_{IR}/\lambda_{V})^{1/4}\simeq 1.8,$ 
where $\lambda_{IR}$ is the (infrared) wavelength of cooling and $\lambda_{V}$
is the (visible) wavelength of heating photons. Since the temperature
varies with distance from the star as $T\propto r^{-1/2}$ (as long as $\tau_{parallel}<1$), 
a cloud of very small particles can be constructed to have color
temperature similar to that of a cloud of larger particles,
as long as it is larger by a factor 
$\sim (\lambda_{IR}/\lambda_{V})^{1/2}\simeq 3$.
Such a cloud of submicron particles would show the silicate feature in 
emission. It's optical depth for starlight propagation 
(constraining the cloud size using the observed color temperature) 
would be
$\tau_{\parallel1}=3.1 M_{18} (\theta_{\frac{1}{2}}/0.8^{\circ})^{-1}$.
Thus a cloud of submicron or few-micron sized particles can account
for the emission, with a modest to small optical depth from
the star to the edge of the cloud and a small optical depth 
emergent from the cloud.
}


It appears therefore that a model cloud that is `physically thick' 
(i.e. thicker than the white dwarf's diameter)
can still be quite flattened, with
angular widths as small as $1^{\circ}$, before becoming optically thick. 
The distinction between the
`physically thick' model \citep{paperone} 
and the `physically thin' model \citep{jura07} is
that the physically thin model is thinner than the stellar diameter, so that
it can be illuminated on its upper and lower surfaces.
In order to intercept as much energy as is observed (3\% of the white dwarf's luminosity), 
a physically thick model at its inner edge must 
have a half-width {\it at least} $10^{9}$ cm, which is larger than the 
stellar radius ($\simeq 8\times 10^{8}$ cm).
Starlight does not illuminate
the surface of the physically thick model, and instead the dust is heated by 
stellar photons that must propagate through the disk. 

\subsubsection{Spherical shell model}

The simplest geometry for the cloud around G29-38, and a limiting case for
the physically thick model, is a spherical
shell. 
A moderately optically thick spherical shell will have the same
temperature distribution as a disk with the same radial density variation.
We use the spherical shell calculations to measure the
radial temperature profile relative to the optically thin case.

\begin{figure}[th]
\epsscale{0.8}
\plotone{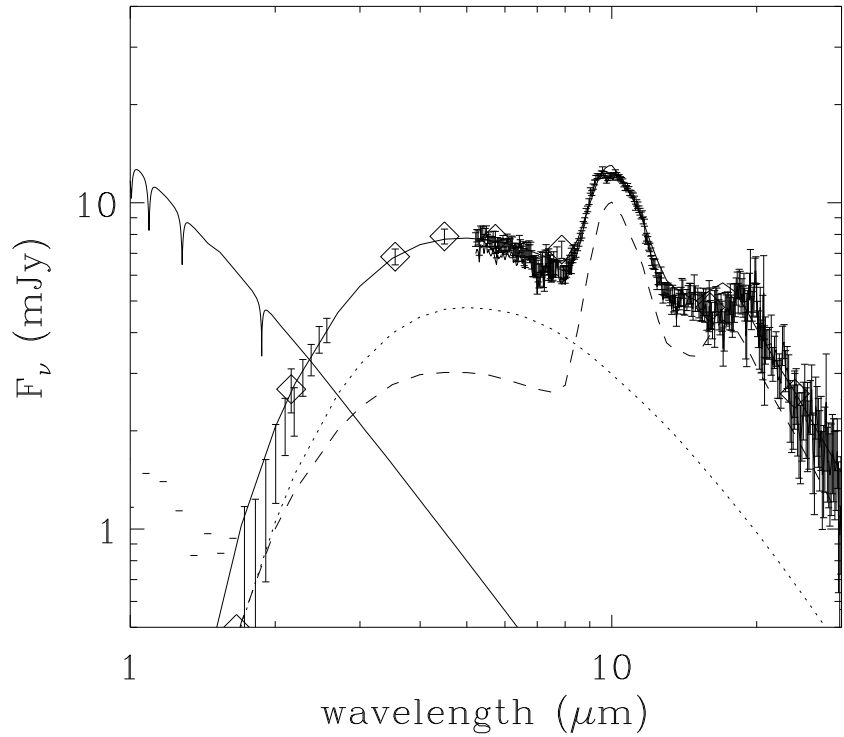}
\epsscale{1}
\caption[g29plotshell]{
Spherical shell models for the dust around G29-38. 
Individual models for amorphous olivine and carbon are shown as dashed and dotted
curves, respectively, and a linear combination is shown as the solid curve.
The input white dwarf spectrum, and the photosphere-subtracted infrared observations, are
shown for comparison. 
The DUSTY models have optical depth $\tau(0.55\mu{\rm m})=0.018$ and 0.011 in silicates and
carbon, respectively.
\label{g29plotshell}}
\end{figure}

Spherical shell calculations were performed using DUSTY
\footnote{Ivezic, Z., Nenkova, M. \& Elitzur, M., 1999, User Manual for DUSTY, University of Kentucky Internal Report, accessible at http://www.pa.uky.edu/$\sim$moshe/dusty.}
for a radial profile $r^{-3}$, 
an amorphous carbon or silicate composition,
and a range of total optical depths, $\tau$ (at 0.55 $\mu$m),
from 0.01 to 10.
The inner boundary of the shell is where the dust temperature reaches 1100 K. 
(Fits with inner temperature 1000 and 1300 yielded significantly worse fits to the
observed spectrum.)
Figure~\ref{g29plotshell} shows emergent spectra compared to the observations.
The first are visually very good, clearly reproducing the spectral shape, despite 
the simplicity of the model;
however, the reduced $\chi^2_\nu=2.2$ is statistically poor. The residuals are dominated by
structure near the 9--11 $\mu$m silicate feature: the observed minus model residuals have
positive peaks at 9.2 $\mu$m and 11.2 $\mu$m and a broad negative trough spanning 8--13 $\mu$m.
The deficiencies of this model are thus due to using only one silicate, in contrast
to the more detailed mineralogy found in the multi-composition models discussed above.

The luminosity of the shell per unit stellar luminosity
is accurately approximated by
\begin{equation}
\frac{L_{d}}{L_{*}}\equiv f \simeq f_{0}(1-e^{-\tau})
\end{equation}
where $f_{0}$=0.74 for carbon and 0.57 for silicate grains.
The observed luminosity ratio, $f=0.03$, could be explained by the combined
carbon and silicate cloud from Figure~\ref{g29plotshell} with optical depth
at 5500 \AA\ and 2000 \AA\ of 0.028 and 0.044, respectively.
The total extinction toward the star is small but could potentially
be measured with precise UV/visible
spectrophotometry, in which case presence of carbonaceous grains could be tested by searching
for a 2175 \AA\ feature such as seen in the interstellar medium \citep{ref2175}.

\begin{figure}[th]
\epsscale{0.8}
\plotone{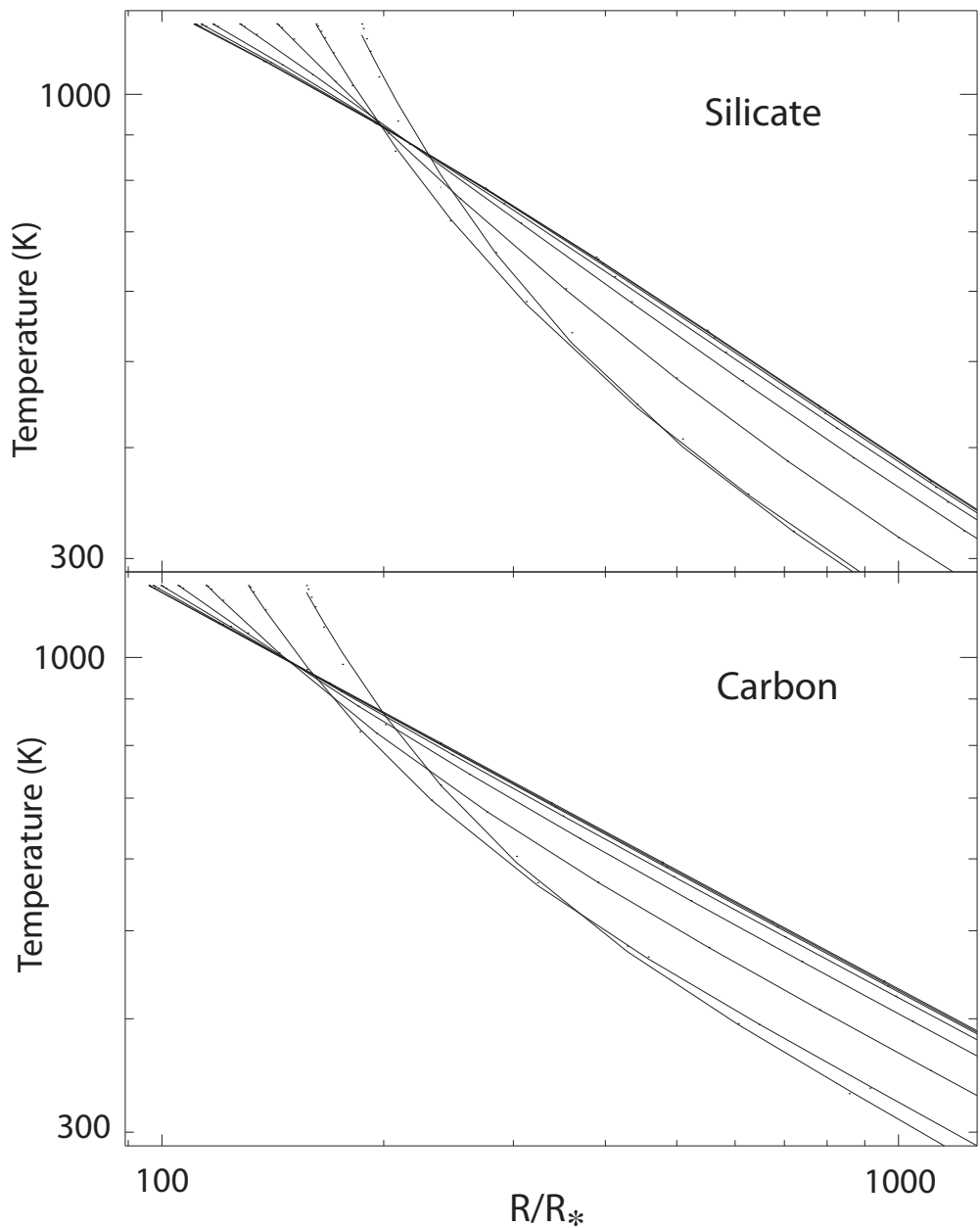}
\epsscale{1}
\caption[g29rad]{
Temperature profiles in a spherical shell around G29-38.
The top panel is for amorphous olivine, 
and the bottom panel is for amorphous carbon.
All models are computed to an inner temperature of 1200 K.
Each profile is for a different total optical depth, with
values 0.0100, 0.0268, 0.072, 0.193, 0.518, 1.39, 3.73, and 10.
The optically thin models extend closest to the star, are the
warmest at the outer edge of the plot, and are closest to a
straight line ($T\propto r^{-0.5}$).
The optically thick models begin further from the star
and decrease in temperature much more steeply.
\label{g29rad}}
\end{figure}

The temperature at a given distance from the
star is the same for a spherical shell as for a moderately optically thick disk,
as long as scattered light and dust thermal emission is a negligible heat source. 
Figure~\ref{g29rad} shows the temperature profiles through shells
with different optical depths.
The temperature versus distance
from the star for grey grains in the optically thin limit
would follow $T\propto r^{-0.5}$. For real materials,
and taking into account radiative transfer, the temperature
profiles are significantly different. 
Using a power-law approximation $T\propto r^{-\delta}$,
the predicted temperature profiles
have $0.44 < \delta < 0.48$ as long as $\tau<0.2$. But
for higher $\tau$ a single power law is not adequate.
We fitted the curves in Figure~\ref{g29rad} 
with empirical functions, for use in the moderately optically thick models.
\def\extra{
Based on this, we found
an approximation formula for the deviation of the temperature
from the greybody value to be
\begin{equation}
\frac{T_{car}}{T_{grey}} = (1+2.5\log\tau+0.818\log\tau^{2})
e^{-r/62.5 \tau^{0.06}} + (0.99-0.236\log\tau+0.082\log\tau^{2})
\end{equation}
for carbon grains, and
\begin{equation}
\frac{T_{sil}}{T_{grey}} = (0.91+2.45\log\tau+1.23\log\tau^{2})
e^{-r/R_{\tau}} + (0.90-0.278\log\tau+0.082\log\tau^{2})
\end{equation}
(where $R_{\tau}=130 R_{*}$ for $\tau<2$ and $R_{\tau}=90 R_{*}$ for
$\tau>2$) for silicates.
Note that these equations were only fitted to 
$0.5 < \tau < 10$ and cannot be extrapolated outside that range. }
For reference, the temperature scalings are closely related to the parameter $\Psi$ 
defined in the original paper on the self-similarity solution used in DUSTY
\citep{ivezic}.

\subsubsection{Moderately optically thick model\label{modthicksec}}

Armed with the temperature profiles, and
the wavelength-dependent cross-sections for each mineral, we can now
compute the brightness of a flattened cloud
that has optical depth $\tau_{\parallel}<10$ and
$\tau_{\perp}<1$. 
We define this as a `moderately optically thick' disk. 
For simplicity we consider a fanned disk, with scale height proportional to 
distance from the star, $h=r\tan\theta_{\frac{1}{2}}$.
The model does not apply to a thin, flat disk or
a disk with a vertical density gradient. 
Full treatment of the optically thick disk with
vertical density gradient requires a
two-dimensional calculation that is beyond our present scope.
\def\extra{
Linear combinations of a flat disk and an optically thin or moderately
optically thick halo can be used to approximate a
disk with vertical density gradient; such hybrid models are
discussed after the individual models are detailed.
}

\begin{figure}[th]
\epsscale{0.8}
\plotone{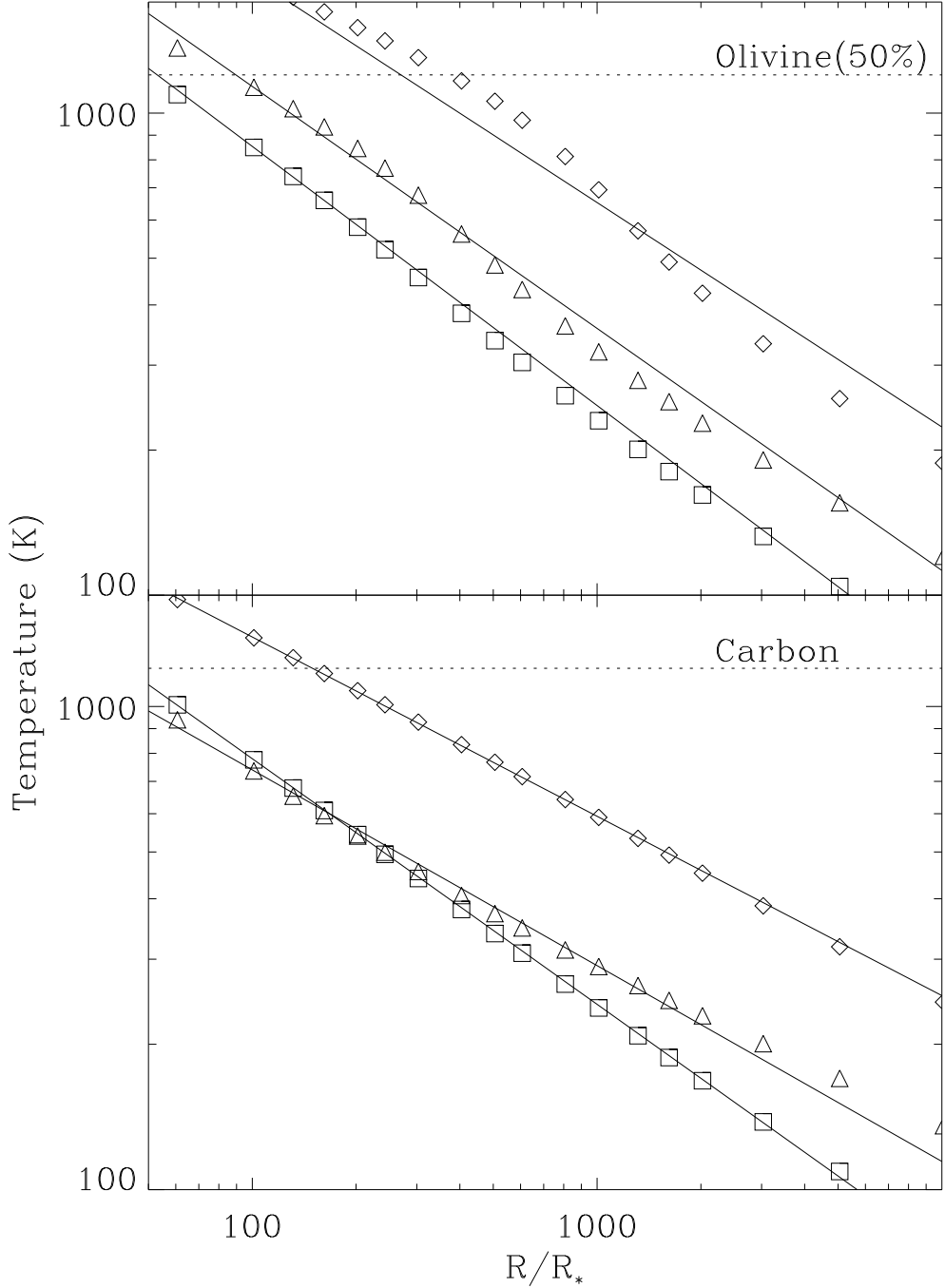}
\epsscale{1}
\caption[tplot]{
Temperatures of olivine (top) and carbon (bottom) grains of three different sizes: 
0.1 $\mu$m (diamonds),
1 $\mu$m (triangles), and
10 $\mu$m (squares). 
The grains are irradiated directly by the white dwarf (i.e.
the cloud is optically thin).
The solid line shows power-law fits of 
the form $T=T_{1} (r/100 R_{*})^{-\delta}$. For olivine (carbon) grains of 
0.1, 1, and 10 $\mu$m radius, 
$T_1=1905$, 1136, 853 K (1393, 740, 780 K) and
$\delta=0.47$, 0.50, 0.54 (0.37, 0.41, 0.51), respectively.
The horizontal dotted line  indicates the
vaporization temperature (1200 K) in our model.
\label{tplot}}
\end{figure}

The moderately optically thick model was calculated for a 
subset of the minerals used in \S\ref{mineralsec}:
amorphous carbon \citep{zubkocar},
amorphous olivine \citep{dorschner},
forsterite 
\citep[pure-Mg crystalline olivine,][]{jaeger03}, and
enstatite \citep[pure-Mg crystalline pyroxene,][]{jaeger98},
and bronzite \citep[Fe-rich crystalline pyroxene,][]{henning97}.
Based on the results of \S\ref{mineralsec}, we expect the main constituents (in order)
to be the amorphous silicates, carbon, and bronzite.
Due to the moderately intensive computations and lack of in-hand UV-FIR laboratory data
(required to span absorption of the white dwarf spectrum as well as thermal emission), 
we did not include as wide a range of minerals as in \S\ref{mineralsec}.
For bronzite, we use the calculations for forsterite but scaled 
them by the relative small-particle emissivity over the range of wavelengths where the optical constants were available 
(6.7--500 $\mu$m).
The most
important compositions for which we did not calculate the moderately-optically-thick model were magnesium-iron sulfides, and water ice (which in fact
cannot exist on grains in this model and requires a separate reservoir, 
as discussed in \S\ref{mineralsec}).
The temperature of grains of each size (from 0.1 to 1000 $\mu$m)
was calculated by balancing radiation from the white dwarf,
geometrically diluted by $r^{-2}$ within $20 < r/R_{*} < 70000$,
with the grain's thermal emission. 
Figure~\ref{tplot} shows the grain temperatures for two materials and three particle sizes in this optically-thin limit.
The temperatures were then adjusted
using the scale factors derived in the previous section appropriate
for the composition, distance from star, and total cloud optical depth. If a grain's temperature exceeds a vaporization temperature, $T_{vap}$, 
it's emissivity is set to zero.
Values of the vaporization temperature in the 1000-2000 K range are expected for most minerals; evidence for inner edges of YSO disks
at distances corresponding to these vaporization temperatures
has been found in interferometric observations \citep{monnier}.
The emission spectrum of dust at each distance from the star
was calculated by integrating over several size distributions: 
a power-law $\propto a^{-3.5}$, a Hanner
law with slope 3.7 and critical size 7.4 $\mu$m,
the size distribution from the coma of comet Halley,
and the size distribution of interplanetary meteoroids
and lunar microcraters \citep{gruen85}.

The emergent spectrum from the cloud was then calculated by 
integrating in spherical coordinates
using a radial density
distribution $n\propto r^{-\alpha}$ and
minimum radius $R_{min}$, with individual models sampled from the ranges
$0.3 < \alpha < 6$ and
$50 < R_{min}/R_{*} < 1000$.
We assume azimuthal symmetry and optically thin infrared emission,
so the opening angle of the disk $\theta_{\frac{1}{2}}$ does not
affect the spectral shape. 
A range of $1000 < T_{vap} < 1600$~K and $0 < \tau_{\parallel} < 10$ were
considered.
Then for each cloud geometry $(\alpha, R_{min})$, a 
linear combination of the models for carbon, olivine, and 
forsterite was fitted to the observations, and the $\chi^{2}$ for
the mixture computed. 
Because these materials vaporize at 
$R<50 R_{*}$ for essentially all particle sizes
(see Fig.~\ref{tplot}), 
all models with $R_{min}<50 R_{*}$ are equivalent.

The best-fitting solution is enumerated in Table~\ref{modthicktab}.
Given 390 data points in the spectral energy distribution,
we expect a `good' fit to have reduced $\chi^{2}_{\nu}\simeq 1$ within dispersion
$\sqrt{2/390}=0.07$. The best-fitting model had $\chi^{2}_{\nu}=1.23$.
We discuss the residuals, which are
localized rather than spanning a wide wavelength range,
in the following paragraph;
for statistical purposes we assume they are unrelated to the cloud geometry. 
The confidence intervals for the parameters, determined by
$\Delta\chi^{2}_{\nu}=0.07$, are listed in Table~\ref{modthicktab}.

The residuals from the radiative transfer model are shown in 
Figure~\ref{fitresid}. Coherent features dominate the residuals; 
the amount by which they increase $\chi^{2}_{\nu}$ measures their statistical significance.
These include the following: 
(1) a `W'-shaped pattern between 8 and 10.5 $\mu$m ($\Delta\chi^{2}_{\nu}=0.30$);
(2) an `S'-shaped pattern between 15 and 21 $\mu$m ($\Delta\chi^{2}_{\nu}=0.24$); and
(3) a `U'-shaped residual between 1 and 3 $\mu$m ($\Delta\chi^{2}_{\nu}=0.20$).
Each of these features is statistically significant.
Feature (1) is due to mismatching the fundamental silicate Si-O stretch feature;
an additional silicate is required which provides a shorter-wavelength silicate feature than
Mg-rich olivine.
Feature (2) is due to either mismatching the silicate Si-O-Si bending feature or presence
of MgO (or Mg$_{x}$Fe$_{1-x}$O with $x>.5$), which has a peak at 18.5 $\mu$m. 
Cosmic abundances favor formation of Mg-rich silicates over Mg oxides, so a preferable
solution for residual features (1) and (2) is a silicate with bluer fundamental and better-matching 
Si-O-Si bending mode features. 

We experimented with the following minerals, inspired by the results of \S\ref{mineralsec}.
All models included amorphous carbon and olivine.
Additional (third) minerals were included one at a time; the most successful 
were bronzite ($\chi^2_\nu=1.22$) and forsterite ($\chi^2_\nu=1.55$), 
in order of goodness of fit. These results are
consistent with those in \S\ref{mineralsec}, where an Fe-rich pyroxene was
shown to be the third-most important mineral.
Then for the most successful third minerals, fourth minerals were added; 
a good combination was forsterite plus montmorillonite ($\chi^2_\nu=1.23$), 
but it was only slightly  better than bronzite model alone. 
Other minerals such as amorphous pyroxene yielded negligible improvement to
the fit ($\Delta\chi^{2}_{\nu}=0.02$). The materials commonly referred to 
as amorphous olivine and amorphous pyroxene are both amorphous silicates and are essentially 
indistinguishable.
It appears to be significant that Fe-rich pyroxene is more abundant
than Mg-rich olivine; this is directly due to
Fe-rich pyroxene having a bluer fundamental Si-O stretch (as observed) compared to
forsterite.
From the laboratory measurements of phyllosilicates by \citet{glotch}, 
there are two
properties that make them possible explanations for residual features (1) and (2): 
phyllosilicates have an
Si-O stretch that is bluer than forsterite, and they have a double-peaked Si-O-Si
bending mode that is similar to the shape observed in the G29-38 spectrum from 18--20 $\mu$m.

The distinction between the bronzite model and the forsterite+montmorillonite model is
mathematically within the errors, but we will refer primarily to the former model
since it is simpler and it agrees with the more extensive mineral search 
from \S\ref{mineralsec}.
The fit could be improved if a more exhaustive mineral search were performed.
However, the results of \S\ref{mineralsec} already demonstrate that including
a wide range of minerals can explain the silicate feature shape adequately,
and we are already at a reduced $\chi^{2}_{\nu}=1.2$ so that the present signal-to-noise
does not allow further, unique modeling.
\def\extra{
Two minerals with significant amplitude in \S\ref{mineralsec} were not included in
this section. Sulfides have a broad bump in the 35--40 $\mu$m region 
(with that of Mg$_{0.5}$Fe$_{0.5}$S at 37.0 $\mu$m and that of FeS [troilite] at 39.2 $\mu$m);
the IRS spectra end at 35 $\mu$m and do not span the red side of that feature, so the presence
or state of sulfides in the disk cannot be uniquely determined.
We also did not include water ice, which was found to be required in the
mineralogical model from \S\ref{mineralsec}. There is still significant H$_2$O in the model
in this section, but it is in the form of water of hydration in the phyllosilicate. 
}

Feature (3) in the residuals relates to the shape of the inner edge of the disk and the vaporization 
temperature. We only explored $T_{vap}$ on a grid with 100 K intervals, and we fixed 
it at the same value for all compositions, so the observed, modest deviations from our model 
are not particularly surprising in the near-infrared. Further, we did not explore the 
composition of the featureless materials, which could affect not only their 
$T_{vap}$ but also the slope of their absorption around 3--6 $\mu$m, which determines the
shape of the inner edge of the disk spectrum.

\subsection{Physically thin disk}

A physically thin
disk is very {\it optically} thick, so starlight cannot propagate radially through the disk.
Such a disk must be so thin that the star can illuminate its surface, or it must be warped such that the surface has clear lines of sight to the star, or a combination of both effects as described by \citet{jura07}.
The temperature versus distance from the star scales as $r^{-.75}$ \citep[][e.g.]{chiang}.
The flux from an optically thick disk is straightforward to
estimate: for G29-38, the disk temperature $T = 8008 (R/R_{*})^{-0.75}$ K,
and the stellar radius $R_{*}=8\times 10^{8}$ cm, 
so the model is determined only by the inner and outer radii
of the disk. If we furthermore set the inner radius to
be that at which dust sublimates, the only free parameter
is the outer radius of the disk. 
Since emission from
the inner radius dominates at the shortest wavelengths,
the optically thick model makes robust predictions of 
the disk flux at the wavelength where emission from
material at the
vaporization temperature peaks, i.e. around 3 $\mu$m.
The flux at longer wavelengths depends on the outer radius.

As observed from Earth, the disk may of course be inclined with
respect to the line of sight \citep{vonhippel}, though previous calculations
considered face-on geometry for illustration
\citep{jura03,jura07}. The flux will scale as $\cos i$, where
$i$ is the angle between the disk axis and the line of sight,
until $i \rightarrow \tan^{-1}H/R_{outer}$, where $R_{outer}$ is the
outer radius of the disk and $H$ is its scale height. Taking the
outer radius $\sim 50 R_{*}$ from the GD 362 
model \citep{jura07}, and requiring disk thickness $H<R_{*}$, 
the low-inclination limit only applies when $i<1.2^{\circ}$.
The nearly edge-on limit is relatively improbable and
furthermore would cover the star unless $H<<R_{*}$, in
which case the edge-on limit applies to even less probable
geometries. 
An optically thick disk has a
spectral energy distribution determined almost entirely
by the outer radius, and total flux scaling with 
$\cos i$.

Using the {\it Spitzer} spectra, it is clear
the simple optically thick model is definitively ruled out
by the presence of a very strong silicate emission feature:
this feature requires an optically thin emitting region.
The feature contributes a significant portion of the disk
luminosity and must have associated continuum. 
Further, the observed spectral energy distribution requires a colder
component with color temperature $\sim 290$ K \citep{paperone} to
explain the flux at 24 $\mu$m.
As a first step toward constraining a possible optically thick disk
around the star, we fitted the thin-disk model to the spectrum excluding
the silicate feature (8--12 $\mu$m), and setting the inner radius as the
location where the grain temperature is 1200 K.  The best fit
has $\chi^{2}_{\nu}=3.8$; this high value is due to the lack of significant 
emergent cold flux from the model.
However we take the constraints on $R_2$ and $i$ as a guide,
with best values $R_{2}/R_{*}=49\pm 5$ and $i=41\pm 3^{\circ}$.

\begin{figure}[th]
\plotone{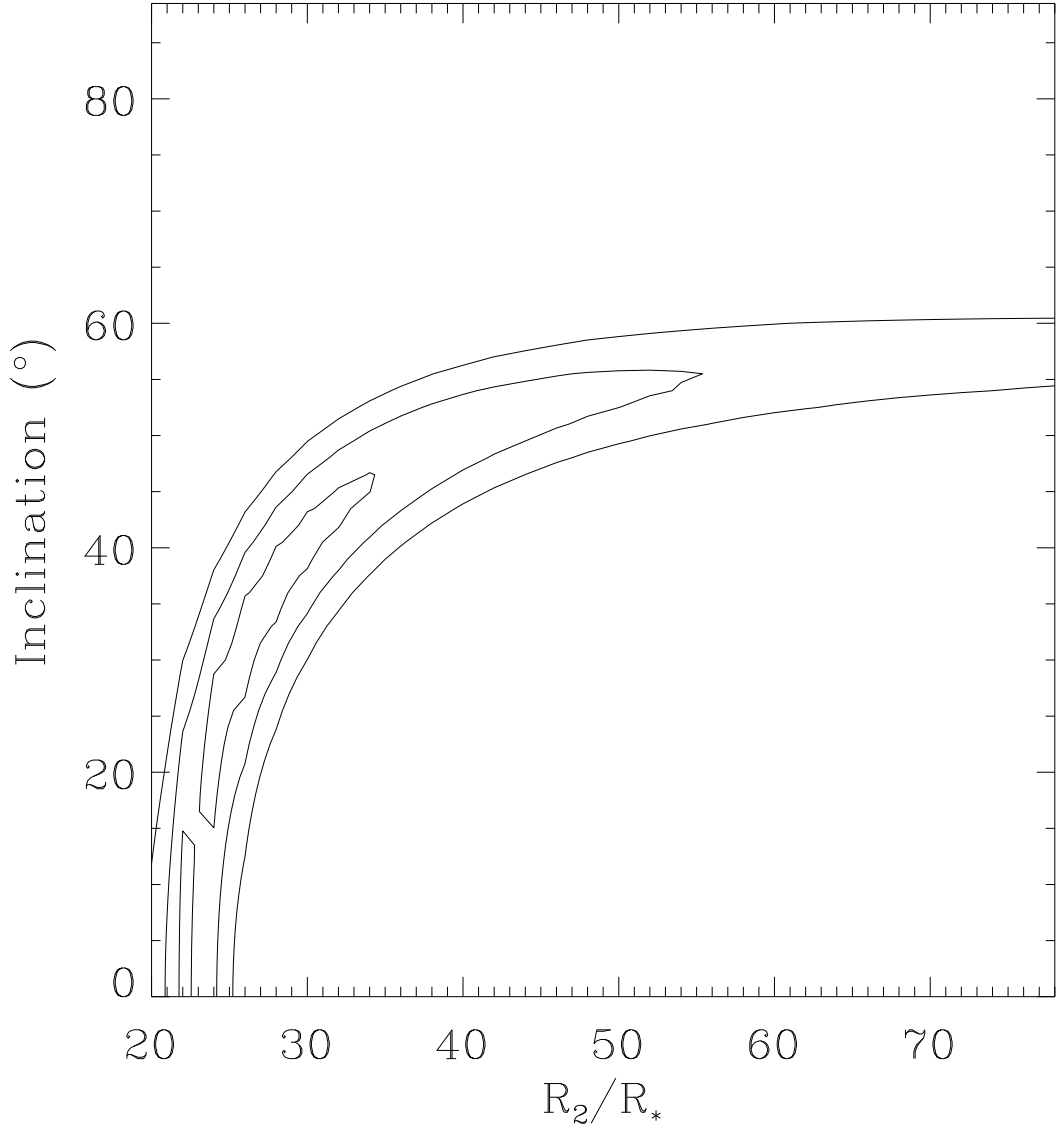}
\caption[juramany]{
Goodness of fit ($\chi^{2}/{\rm d.o.f}$) from the physically-thin, 
optically-thick model for a range of outer radii ($R_{2}$, in units of stellar
radius) and inclination. The best fitting model has inner and outer radii $R_{1}=9$,
$R_{2}=25$, and inclination $23^{\circ}$. 
\label{juramany}}
\end{figure}

Instead of attempting to fit the entire spectrum with the thin-disk model, we
now consider only fitting it to the shorter-wavelength continuum, with an
eye toward adding a cooler, physically-thick silicate-bearing cloud.
Thus we excluded wavelengths longer than  8 $\mu$m and 
fitted the physically thin model to the spectrum of G29-38.
The near-infrared spectrum, and the decrease from 5--8 $\mu$m, require
the inner edge of the disk is closer to the star than the point that reaches
1200 K; a better fit is obtained with $T_{vap}=1500$ K so that $R_{1}=9R_{*}$.
Figure~\ref{juramany} shows the constraints on outer radius ($R_{2}$) 
and inclination ($i$). 
The best fit has a low $\chi^{2}_{\nu}=0.8$; it is probably less than 1 due
to overestimation of the uncertainties due to photospheric subtraction in
the near-infrared.
The best fitting outer radius $R_{2}=25R_{*}$ and inclination $23^{\circ}$.
It is notable that the required inclination is roughly in the range
required by \citet{graham} in his model
for the near-infrared timing (which required an inclination
such that the dust temperature pulsations are detectable
while the exciting pulsations on the photosphere are not).

An amendment to the physically thin disk model is needed to
improve the fit and explain the silicate emission.
One possible solution is to include
an optically thin, effectively-physically-thick region at 
the outer edges of the ring. \citet{jura07} showed that
for GD 362, an extension of the ring that is warped, by only
$\sim 7^{\circ}$, can produce a silicate feature in emission.
Figure~\ref{cartoon} shows their model, compared to the moderately thick model
from \S\ref{modthicksec}.
Indeed this outer region produces both the silicate feature
and a significant fraction of the continuum at 
$\lambda>11$ $\mu$m in their model. The plausibility of
the warp of the outer disk is discussed and justified physically by
\citet{jura07}, who label it as `region III' in their model. 
In many ways, the details 
(mass, temperature, radius, vertical extent)
of this outer
portion of the disk must be similar to the physically
thick model, since they explain the mid-infrared emission
in the same way (an optically thin cloud of silicates). In the
warped disk model, the optically thin region is actually the
upper layer of the disk; i.e. even the warped portion of
the disk may be optically thick, as long as it is has a direct
line of sight to the star.

\subsection{Comparing the thick and thin models}

The disk spectral energy distribution can be empirically 
decomposed into three major components. One of them is
a  continuum with a relatively hot (890 K) color temperature, peaking
around 4 $\mu$m and dominating the near-infrared emission.
Another component is continuum with a lower color temperature ($\sim 300$ K).
The other major component is the silicate emission feature.
In the \citet{jura07} model, the hot component is the
blackbody disk---hot because it is close to the star---and the
cooler continuum and silicate feature arise in the outer warp region III.
In the physically thick model (in Paper I and \S\ref{modthicksec} above), 
the hot component is
amorphous carbon---hot because it is due to highly absorbing
material---and the cooler component is silicates---cool because the
silicates are more transparent and have strong mid-infrared emission
features that allow them to cool efficiently;
the two materials are colocated.

For many plausible configurations of solid material around
the white dwarf, we can consider the cloud as the sum of an
optically thick disk and an optically thin halo or flared disk surface. 
 Indeed, \citet{vinkovic} proved 
mathematically that flared disk models are equivalent
to disk plus halo models. The radial
profile of a spherical halo can be directly related
to the flaring angle of a flared disk. Thus it is not
possible, using the spectral energy distribution alone,
to separate disk and halo (or warped disk) models.
The halo dominates the infrared emission when 
$\tau_{halo}>H/4R$, where $H$ is the flare height and
$R$ the distance from the star \citep{vinkovic}. 
For the wedge-shaped 
`physically thick' model discussed above, this is
equivalent to $\tau_{\parallel}>\tan\theta_{\frac{1}{2}}/4$. 
Equating the emission and absorbed flux (3\% of the stellar
flux), the halo will dominate the infrared emission
when $\tau_{\parallel}>0.004$, which is already required
for both the physically thick and thin models.

\begin{figure}[th]
\plotone{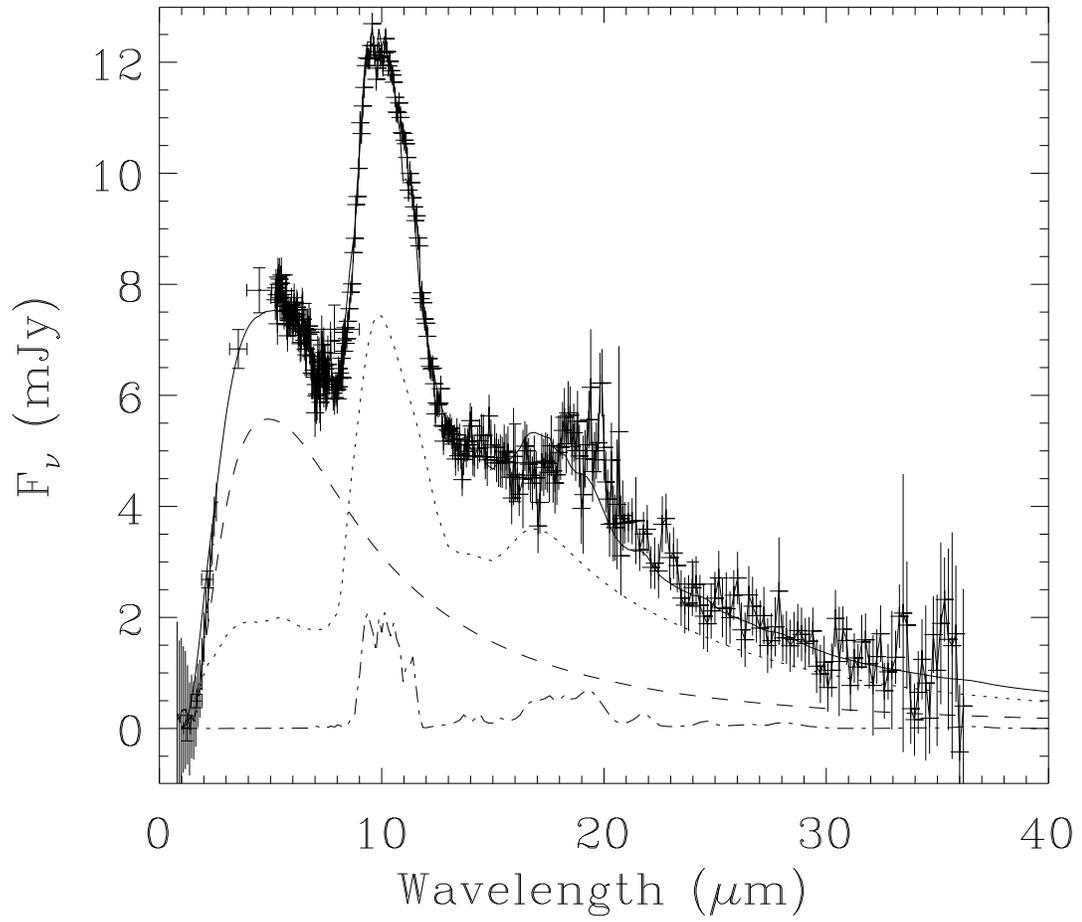}
\caption[fplotbest]{
Best-fitting moderately-optically thick disk model (solid line)
combining amorphous carbon (dashed), amorphous olivine (dotted),
crystalline bronzite (dash-dot)
to the observed photosphere-subtracted spectral energy distribution of G29-38. 
\label{fplotbest}}
\end{figure}

\begin{figure}[th]
\plotone{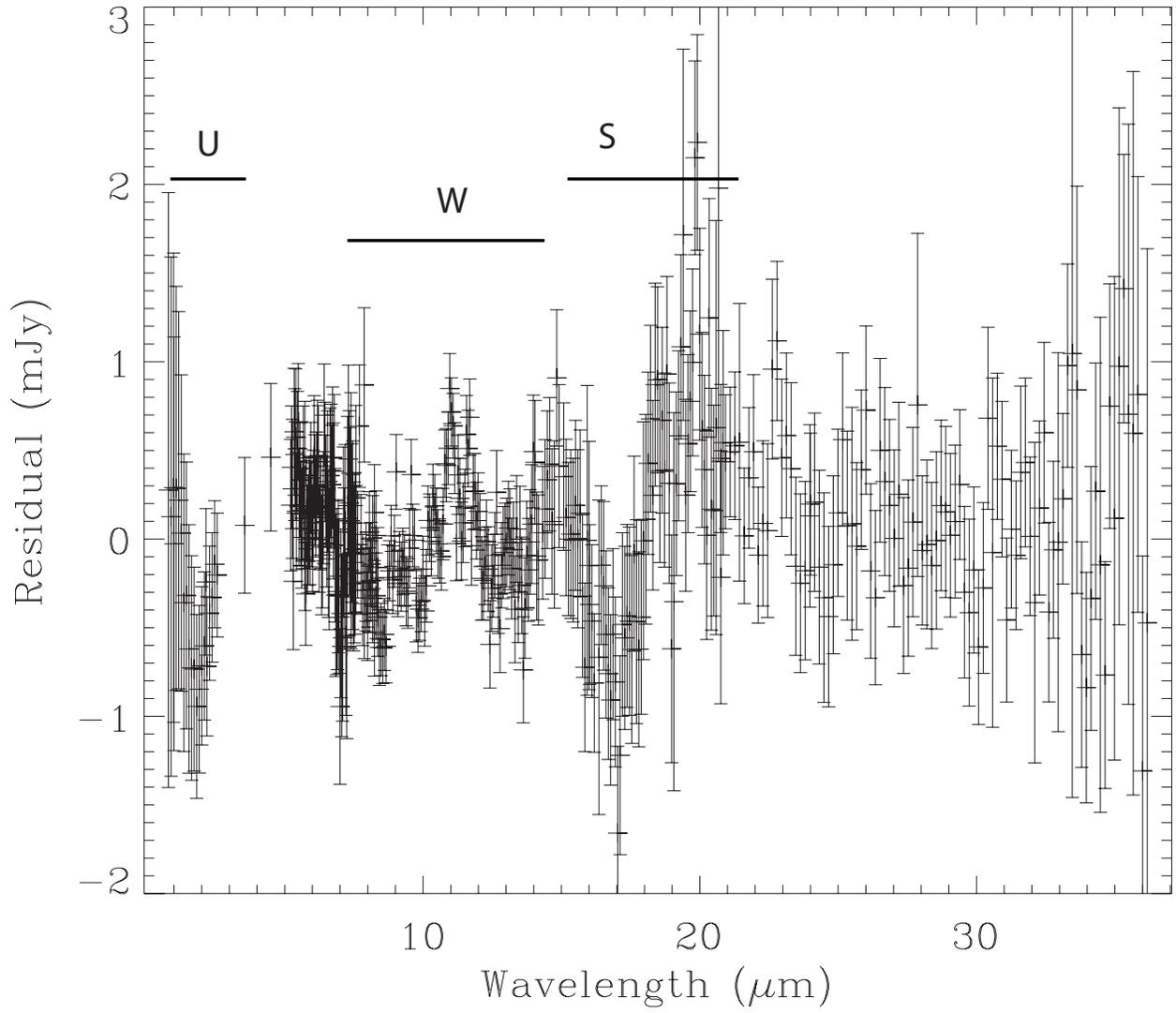}
\caption[fitresid]{
Residuals (observed minus model) from the best-fitting 
moderately-optically thick disk model of Fig.~\ref{fplotbest}.
Features discussed in the text are labeled.
\label{fitresid}}
\end{figure}

\begin{figure}[th]
\plotone{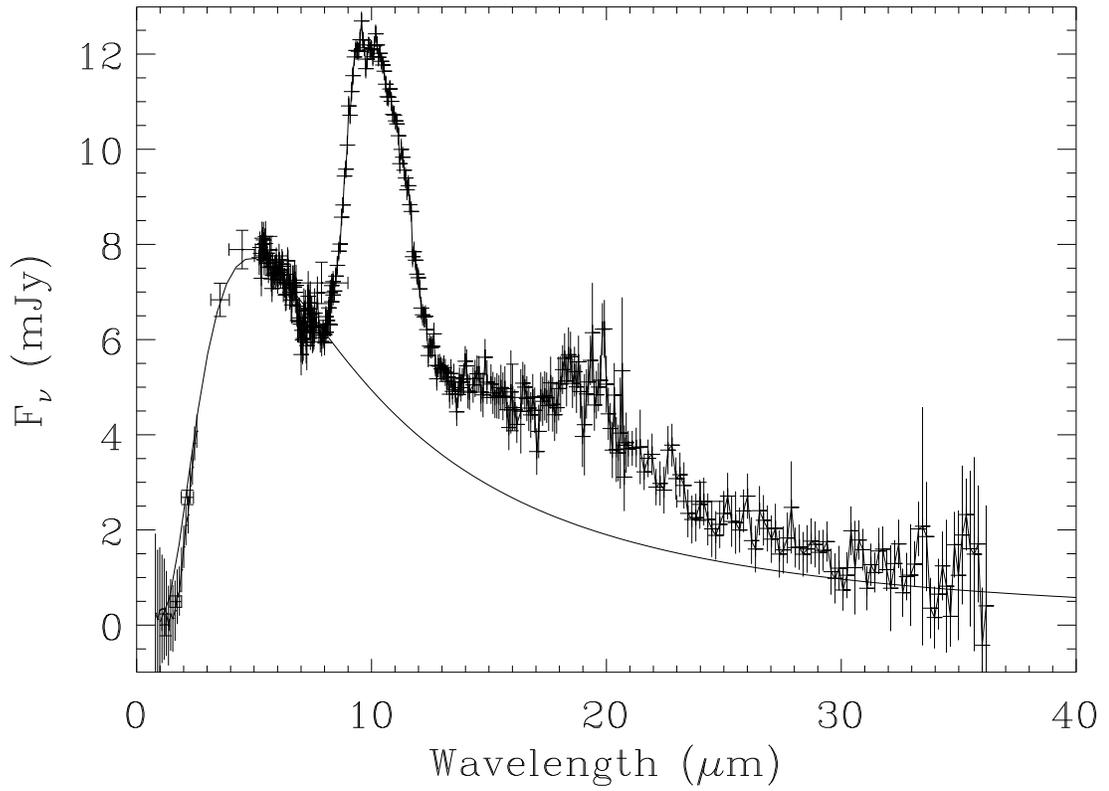}
\caption[juramany]{
Best-fitting physically-thin, optically-thick disk model. The model (solid line)
was only fitted to data at wavelengths less than 8 $\mu$m;
it cannot produce the silicate feature and under-predicts
(by far) the longer-wavelength continuum.
\label{jurabest}}
\end{figure}

\begin{figure}[th]
\plotone{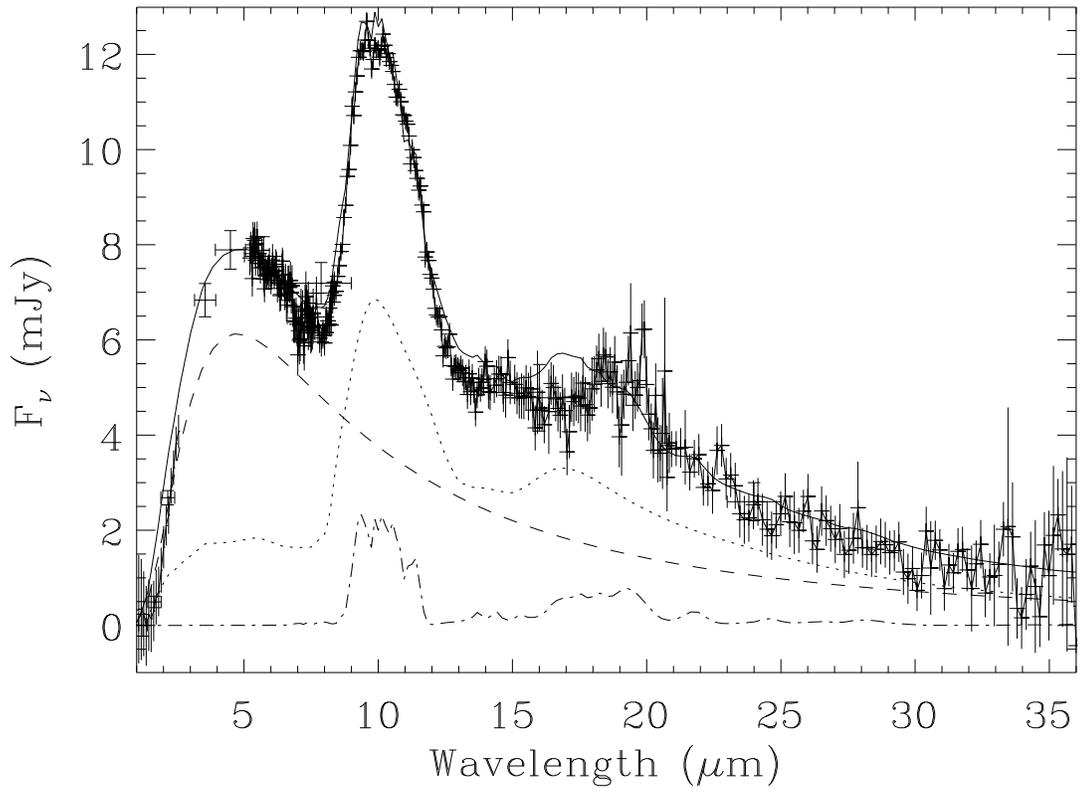}
\caption[thinthick]{
Best-fitting combined model, with a physically-thin, optically-thick disk 
(cf. Fig~\ref{jurabest}) plus a
silicate-only (no C or Fe) physically-thick disk (cf. Fig.~\ref{fplotbest}).
\label{thinthick}}
\end{figure}

For the specific case of the G29-38 disk, we made a direct substitution of the
optically-thick disk for the C/Fe component of the moderately-optically-thick model.
The optically-thick disk was taken directly from the model in Figure~\ref{jurabest} (i.e.
the one fitted to wavelengths shorter than 8 $\mu$m), and the silicate components
were taken from the moderately optically thick model in Figure~\ref{fplotbest}. 
The silicate component of the moderately-optically-thick disk was rescaled in 
amplitude in order to match the observed flux
after being added to the optically thick disk. 
The parameters for the best-fitting disk in this combined disk+silicate model are
similar to those derived above using only short-wavelength data; $R_{2}=22R_{*}$ and
$i=29^{\circ}$.
Figure~\ref{thinthick} shows the best fit;
it is very similar to the moderately optically thick model in Figure~\ref{fplotbest}; 
the goodness-of-fit, $\chi^{2}_{\nu}=1.40$, not much worse than 
the moderately optically thick model. Improvements to $\chi^{2}_{\nu}$ as obtained
in \S\ref{mineralsec} could be obtained by including more minerals in the optically thin
region.

\section{Timing constraints\label{varsec}}

The geometry of the G29-38 disk can been constrained 
 using timing information.
The star is a ZZ Ceti variable with non-radial modes that yield
its optical pulsations. 
\citet{graham} found
pulsations at periods $\sim 200$ sec in the K and L bands that have 
no strong counterpart in J and B-bands.
More detailed optical photometry clearly shows pulsations at
the predicted frequency: see the peak around 5380 $\mu$Hz
from the Whole Earth Telescope observations
\citep{winget}. The weakness of this mode at visible wavelengths,
compared to infrared wavelengths, could be due to the manner in 
which it interacts with the disk.
The B-band light is completely dominated by direct photons from
the star, while the K-band light contains a contribution from
thermal emission by the dust that produces the mid-infrared excess.
The K-band pulsations are attributed to dust temperature variations.

As \citet{graham} explained,
to generate observable K and L-band pulsations from a mode that is
not prominent on the photosphere,
the geometry is constrained. 
They invoke a mode of
stellar brightness variations (2$^{nd}$ order spherical harmonic) that
brightens the pole and dims the equator of the star.
If the disk is thin enough that its heating is driven by the equatorial
stellar brightness, then it will track the equatorial photospheric
temperature, rather than the average over the surface.
\citet{graham} show that the spherical harmonic that excites the
disk temperature variations can exist without significant variation
of optical wavelength light if the star is viewed at the
angle where the ratio of the two spherical harmonics is small.
To eliminate J-band variability in their data required a viewing 
angle in the range 45--65$^{\circ}$. 

The IRAC observations confirm the significance of the 200 sec period
in the near-infrared. (No modes with periods longer than 500 s are detectable
due to the brevity of the IRAC observations, so
we could not address the 615 sec period that dominates the optical pulsations \citep{winget}.)
The fluctuating portion of the infrared flux, after
subtracting the direct stellar photosphere, is 7\% at 2.2 $\mu$m,
5\% at 3.6 $\mu$m, and $<3$\% at 8 $\mu$m. The color of the
fluctuations suggests they arise from the inner portion of the
disk, near the dust vaporization temperature.
This explanation works well with the physically thin disk model
with the inner edge of the disk at 1200 K and decreasing temperature
outwards.
The timing constraints rule out a spherical distribution for all 
the dust, though some dust can remain in a spherical distribution
without generating pulsations (because the temperature variations
average out over a sphere).
The observations do not yet rule out a
`physically thick' configuration with a small opening angle ($\theta_{\frac{1}{2}}$).
We consider the \citet{graham} results very
important; they require observational confirmation
and have the promise of revealing the disk structure in
more detail.

\section{Discussion and Conclusions\label{discussion}}

The mid-infrared spectrum of G29-38 is due to a cloud of small particles orbiting the
star within the distance where tidal forces from white dwarf's strong gravity
would destroy a large, weak body. The tidal forces exceed self-gravity and
strength of a non-rotating, rigid spherical body within a distance
\begin{equation}
d/R_{*} = 120 \rho^{-1/3} \left[ 1 + 0.11 \frac{S_{\rm kPa}}{\rho r_{10}^{-2}} \right]
\label{roche}\end{equation}
where $\rho$ is the bodies density (g~cm$^{-3}$), $r_{10}$ is its radius (in units of
10 km), and $S_{\rm kPa}$ is its strength in kPa. 
If the dust around G29-38 was produced by one body, then the present-day mass 
requires $r_{10}>1$. 
For an asteroid or comet, we expect $0.5<\rho<2$. 
Spin rates of near-Earth objects require strengths $\sim 2 r_{10}^{-1/2}$ kPa
to balance centrifugal forces against strength \citep{holsapple}.
The strength term in equation~\ref{roche} is thus negligible for
bodies with $r_{10}\rho>1$, which is likely for the parent bodies of the
G29-38 dust. 
Fragments of parent bodies can survive much closer to the star;
rocks with $S_{kpa}=3000$, $\rho=3$, and could survive against tidal
disruption all the way to the
surface of the star if they are smaller than 0.1 km.

In the spectrum of G29-38, the strong emission feature at 9--11 $\mu$m, the color temperature, and the timing information all support the interpretation of the mid-infrared excess as being dominated by a cloud of small silicate particles. The infrared excess at 2--6 $\mu$m is apparently due to a featureless blackbody continuum. Based on our modeling results, this higher-color-temperature emission could be explained either by a highly-absorbing mineral (like solid C or Fe) or by a massive disk of material that is thinner than the white dwarf's diameter. 
The difference between these models, both of which can explain the observed spectral energy distribution reasonably well, is significant: the massive disk model could harbor 
$10^{24}$ g (or more) of dust \citep{jura03}, while the cloud of amorphous C or Fe requires $\sim 10^{19}$ g. There is no {\it a priori} reason that either of these mass estimates should be preferred or rejected. Some $10^{19}$ g of silicates are required in all models. 

To understand the origin of the circumstellar material, it is important
to know how much mass is observed. If only $10^{19}$ g of material is present, then the observed infrared excess can be explained by a single small asteroid or a comet. 
If $10^{24}$ g of material is present, then an entire, large asteroid, or numerous
smaller ones, is required. In all cases, the parent body must have been somehow transported from a distance far enough from the star that it could have survived the red giant and AGB phase of the star ($>5\times 10^{13}$ cm) inward to the Roche limit where it would be disrupted by the gravity of the white dwarf ($< 10^{11}$ cm). 

\subsection{Nature of the parent bodies}
The mineralogical results can help us relate the dust to possible parent bodies.
We concentrate on the two primary components of the spectrum separately: the silicate material (required in all models in order to explain the 9--11 $\mu$m emission feature) and the highly-absorbing material (required only in the thick-disk models).

The composition  of the highly-absorbing material cannot be determined unambiguously from the spectrum. We fitted it in \S\ref{mineralsec} with amorphous C, based on the high cosmic abundance of C. But solid Fe or Si are also plausible, given their high cosmic abundance. 
Indeed, for chondritic (asteroidal and terrestrial) material the abundance of Si and Fe is
much higher than C \citep{juracarbon}.
Mineralogical models for dust around other stars and in comets, using the same methodology as in \S\ref{mineralsec}, do not always show a high abundance of C 
\citep{lisse07a,lisse07b}. For extrasolar systems, a careful subtraction of the photosphere is critical to measuring the high-color-temperature emission (which is characteristic of highly-absorbing material like solid C). 
For G29-38, the high-color-temperature excess (at 3.6 $\mu$m and
longer wavelengths) is so far above photosphere that it must arise from 
circumstellar material, but
at wavelengths shorter than 2 $\mu$m the infrared excess shape depends on the photosphere model. 
For G29-38 the mid-infrared data clearly require emission with a color temperature $> 800$ K, whether it is C or Fe.

The compositions of the potential parent bodies for the circumstellar material around white dwarfs can be addressed by studies of Solar System bodies.
To date there has been no sample return mission from an asteroid, but meteorites provide direct measures of composition of parts of at least some asteroids. Carbonaceous chondrites 
have some C, but all chondrites are largely silicate mineral (olivine and pyroxene, 
mostly Mg-rich [forsterite and enstatite]), 
with a wide range of other minerals (some Ca and Al-rich)
and metals (often including previously-molten Fe).
Metallic meteorites, commonly found in museums and on the ground, have largely Fe and Ni composition \citep{shearer}.
Comets are likely to have a more primitive composition than carbonaceous chondrites, with 
abundant silicate grains as well as carbonaceous material, based on infrared spectroscopy of cometary dust, laboratory study of cometary interplanetary dust particles, and {\it in situ} mass spectrometry during the 1P/Halley flyby in 1986 \citep{hannerbradley}. 
The most abundant silicate minerals in meteorites are Mg-rich olivines and pyroxenes, 
as well as feldspar and phyllosilicates. Interplanetary dust particles believed
to originate from comets (CP type) are largely composed of phyllosilicates that
require aqueous alteration on their parent body \citep{messenger}.

Based on the analogy to Solar System bodies, the dominance of Fe-rich pyroxene mineralogy is
distinct. Comets or outer-main-belt (D-type) asteroids contain Mg-rich olivine and phyllosilicates,
which when combined can reasonably fit the observed spectrum of G29-38.
Both comets and D-type asteroids contain organic material, which would be consistent
with the presence of amorphous C, and the most primitive carbonaceous (CI) 
meteorites are largely composed of phyllosilicates.
But the Mg-rich olivine plus phyllosilicate model is not as good a fit to the data as Fe-rich
pyroxene. There are pyroxene-dominated meteorites, but they are dominated by Mg-rich pyroxene
(hence the name enstatite meteorites for this rare class).
It is worth noting that enstatite chondrites are thought to have formed in reducing conditions and contain
other minerals including niningerite \citep{weisberg}, which was one of the most abundant minerals from
our fit for G29-38 in Tab.~\ref{minfittab}. Thus the physical conditions for formation of the
enstatite meteorite parent bodies may have some relevance to the formation of the G29-38 debris
parent body.

In terms of the featureless material that produces the near-infrared continuum (in the physically
thick models), either C or Fe are acceptable to the fits. If the material were C, then a
cometary or D-type asteroid origin would be more likely, whereas with abundant Fe, formation
closer to the star and within a differentiated parent body would be implicated.

Some constraints on the composition of the material are obtained from the abundances in the white dwarf atmosphere. 
Solid material at the inner edge of the disk is constantly being vaporized by the stellar radiation. These vapors reach the atmosphere of the star (or are blown out of the system). 
They cannot reside in the stellar atmosphere for long; instead, they diffuse 
rapidly inward, deeper than the photosphere. 
Thus the heavy elements in the stellar atmosphere must be `fresh,'
consistent with an origin from vaporization of circumstellar dust but not with a long-lived stellar atmosphere. 
G29-38 has metals present in its atmosphere.
\citet{juracarbon} discussed the deficiency of C in some externally polluted
white dwarfs, where the abundance of C relative to Fe is more than
10 times lower than solar. 
CI meteorites have C/Fe 10 times lower than solar; this is commonly
explained by the volatility of C and the high temperature required for chondrite formation
\citep{brearley}.
We note that the abundances of refractory elements in comets
{\it and} asteroids as well as that inferred from exozodiacal dust
is consistently less than solar, in the 7--10 \% range 
\citep{lisse06}. But in a relative abundance, dust collected during the encounter 
with the long-period comet 1P/Halley has
C/Fe abundance ratio similar to Solar \citep{halley}. 
Short-period comets (which are periodically heated to higher temperature) may be more devolatilized and may have abundances more similar to asteroidal material. 


\subsection{Disruption of `Hot Jupiter'?}
One intriguing possibility for the origin of the infrared excess is the survival of the core of a giant planet and its subsequent gravitational disruption. 
`Hot Jupiters,' with masses of order
$10^{30}$ g orbiting within $10^{12}$ cm (0.1 AU) of their star,
appear to be fairly common in extrasolar planetary systems: 1.2\% of nearby F, G, and K stars
has such a planet \citep{hotjupiter}.
In this scenario, the planet would become engulfed into a common atmosphere during the red giant phase. Drag from the extended stellar atmosphere would cause the planet to spiral inward toward the stellar core. At the end of the mass-losing phase of the star's evolution, we would be left with a white dwarf, the surviving core of the planet, and the planetary nebula composed of the outer atmosphere from the star. 

If a planet began at $<10^{12}$ cm from the star, it is possible
that its remnant could land within the Roche radius. 
The effect of the post-main sequence evolution on the planetary dynamics has not been explored in detail.
The change in mass of the central star could make any borderline-unstable system of
multiple planets unstable \citep{debes02} and could lead to nonlinear orbital perturbations.
A simple estimate of the orbital decay due to gas drag when the planet is within the red
giant atmosphere is made by setting the rate of kinetic energy imparted to the planet,
\begin{equation}
\dot{E} = \frac{1}{2} \rho \pi R_p^2 \left(\frac{GM_*}{a}\right)^{3/2}
\end{equation}
equal to the change in orbital binding energy
\begin{equation}
\dot{U} = \frac{G M_* M_p}{a^2} \frac{{\rm d}a}{{\rm d}t},
\end{equation}
where $M_p$ and $R_p$ are the mass and radius of the planet, $G$ is the gravitational constant,
$a$ is semimajor axis of the presumed-circular orbit, 
$M_*$ is the mass of the star, and $\rho$ is the mass density of
the star at the distance of the planet. Taking for illustration a planet with the mass of Jupiter
and average density 1 g~cm$^{-3}$,
orbiting at a distance of 0.1 AU from a star of mass $2 M_{\odot}$ that has mass density
$10^{-9}$ g~cm$^{-3}$ at 0.1 AU from its center, the timescale for orbital decay 
is $6\times 10^{4}$ yr. This is much smaller than the asymptotic giant branch lifetime 
$\sim 10^{7}$ yr \citep{vassiliadis} for a progenitor star mass of $3.1 M_{\odot}$ \citep{weidemann}.
Thus the inward spiral of hot Jupiter (or other close-in) planetary orbits is likely to be rapid, once
the star has entered its giant phase.


Considering only a convective main-sequence primary star, \citet{sasselov} showed
that the orbit of a `hot Jupiter' like OGLE-TR-56b would tidally decay on a timescale of 1--10 Gyr.
Again considering only a main sequence primary star, 
\citet{baraffe} showed that planets with masses less than
$3\times 10^{30}$ will evaporate down to a rocky core in less than 5 Gyr
These effects may combine, with tidal decay enhancing evaporation \citep{erkaev}.
Tidal decay of `hot Jupiter' orbits cannot be too rapid for main sequence stars, or 
else they would be much less common. 
The orbital evolution is very sensitive to 
the stellar structure and the planet's orbital eccentricity; the response is
highly nonlinear and may lead to a rapid inward spiral upon perturbation of
the stellar interior structure or eccentricity \citep{ogilvie,jackson08}
Nonetheless, regardless of whether tidal decay can bring planets close to the stellar
surface during main sequence evolution, the gas drag during the red giant phase 
should finish the job.

How much of a `hot jupiter' could survive the late stages of its star's evolution is not known.
A brown dwarf was recently discovered around a white dwarf suggesting that some companions can 
survive the red giant phase of the primary star despite being well within the atmosphere
\citep{maxted}. Indeed, the companion is within the Roche distance of the white dwarf
unless it has a mass density greater than 3.6 g~cm$^{-3}$ or is held together by more than
its own gravity.
The composition of the core of an extrasolar giant planet is  difficult to constrain. Mass and radius estimates seem to require a dense core, possibly of high-pressure ice or stony composition 
\citep{burrows07}. This material would be tidally disrupted, as in
the asteroid (or comet) disruption models considered to date. 
The Fe-rich pyroxene mineralogy for the G29-38 debris, which is distinct from that
of most comets and asteroids, does not appear impossible for a rocky planetary core, but
at present there is little more than can be said about whether such a mineralogy is likely.

That G29-38 is exceptional
among debris disks (having by far the brightest infrared excess, with 3\% of the star's
luminosity absorbed and radiated by dust)
makes the hypothesis more attractive. Consider the reverse argument: if
1\% of all stars have `hot jupiters,' what is their expected end state? From this point
of view, it seems inevitable that such remnant planets would generate debris for their
parent star in its white dwarf phase. But this argument assumes that `hot jupiters' 
exist around A-type stars (the progenitor type for G29-38), at least part of the planet can
survive the common envelope phase, the remnant core lands within the Roche radius, and
the debris can survive 500 Myr (the age of G29-38). 
The age problem may not be severe, if 
the remnant planetary core becomes fragmented, with fragments gradually entering the
Roche radius due to collisional disruption and gravitational perturbations.
Further theoretical work is needed to test the viability of this scenario.

\acknowledgements  

WTR gratefully acknowledges discussions on radiative transfer in disks with Moshe Elitzur in May 2007.
WTR gratefully acknowledges discussions on extrasolar giant planets with Adam Burrows in June 2007.
This work is based in part on observations made with the {\it Spitzer Space
Telescope}, which is operated by the Jet Propulsion Laboratory, California
Institute of Technology under NASA contract 1407. 


\clearpage
\begin{table}
\caption[]{{\it Spitzer} Observing log for G29-38}\label{obstab} 
\begin{flushleft} 
\begin{tabular}{llll}
Date & AORID & Instrument & Wavelengths \\ 
\hline
2004 Nov 26 & 10119424 & IRAC & 4.5, 8 $\mu$m \\
2005 Nov 26 & 11124224 & IRAC & 3.6, 4.5, 5.8, 8 $\mu$m \\
2004 Dec 2  & 10149376 & MIPS & 24 $\mu$m \\
2004 Dec 8  & 10184192 & IRS & 5.2--14.2 $\mu$m\\
2005 Dec 23 & 13835264 & IRAC & 3.6, 4.5, 5.8, 8 $\mu$m \\
2006 Jun 30 & 13828096 & IRS & 5.2--36 $\mu$m \\
\end{tabular}
\end{flushleft} 
\end{table}  

\clearpage

\begin{table}
\caption[]{Mid-infrared Fluxes of G29-39 (mJy)$^a$}\label{iractab} 
\begin{flushleft} 
\begin{tabular}{lccc}
	&2004 Nov 26 & 2004 Nov 26 & 2005 Dec 23\\
    & 10:54     & 10:58       & 23:23\\
\hline
3.6 $\mu$m &                & $8.37\pm 0.01$ & $8.10\pm 0.03$ \\
4.5 $\mu$m & $8.88\pm 0.02$ & $8.87\pm 0.01$ &    \\
5.8 $\mu$m &                & $8.37\pm 0.02$ & $8.28\pm 0.07$\\
8.0 $\mu$m & $8.73\pm 0.03$ & $8.72\pm 0.02$ &     \\
\end{tabular}
\noindent\par $^a${Uncertainties are statistical uncertainty
in the weighted mean of the flux measurements from all frames
taken during each observing sequence}
\end{flushleft} 
\end{table}  

\clearpage

\begin{deluxetable}{lcccccc}
\tablecaption{Composition of the Best-Fit Model\tablenotemark{a} to the {\it Spitzer}/IRS G29-38 Spectrum\label{minfittab}}
\tabletypesize{\footnotesize}
\tablehead{
\colhead{Species} & \colhead{Weighted\tablenotemark{b}} & \colhead{Density}&  
\colhead{Mol. Wt.} & \colhead{$N_{moles}$\tablenotemark{c}} & 
\colhead{$T_{max}^{d}$}& 
\colhead{$\chi^{2}$ if}\\
& Surface Area & (g~cm$^{-3}$) & (relative) && (K) & excluded 
}
\startdata
\cutinhead{Detections	}
Amorph Olivine (MgFeSiO$_4$) & 	0.33	 &	3.6	&      172	&	0.69	&   890	&	90.6\\
Fayalite (Fe$_2$SiO$_4$)		&	0.08	&	4.3&	      204&		0.17&	   890&		2.91\\

FerroSilite (Fe$_2$Si$_2$O$_6$)&  0.11&		4.0&	      264&		0.17&	   890&	9.85\\
Diopside (CaMgSi$_2$O$_6$)	 &    	0.05	 &	3.3	&     216&		0.076&	   890&		2.05\\
OrthoEnstatite (Mg$_2$Si$_2$O$_6$)&  	0.04	&	3.2	&      200&	0.064&	   890&		1.98\\

Niningerite (Mg$_{10}$Fe$_{90}$S)\tablenotemark{e} &0.10&4.5& 84&	0.53&  890&		1.49\\

Amorph Carbon (C)&      	0.28&		2.5&       12&		5.83&	   930&	$> 100$\\

Water-ice (H$_2$O)&     	0.29&		1.0	&     18&	1.61&	   220&		5.82\\
						
\cutinhead{Upper Limits and Non-Detections}

Forsterite[Koike](Mg$_2$SiO$_4$)&  	0.02&3.2&      140&	0.046&   890&	1.15\\
Amorph Pyroxene (MgFeSi$_2$O$_6$)&  	0.00	&3.5&      232&	0.09&   890&		1.04\\

Smectite/Notronite\tablenotemark{f} &     	0.00	&	2.3&	    496&		0.03&	   890&		1.04\\

Water Gas (H$_2$O)&    	0.01	&	1.0	&       18&	$\leq 0.00$&   220&		1.04\\

Magnesite (MgCO$_3$)&  	0.00&	3.1&     84&	$\leq 0.00$&	   890&		1.04\\
Siderite (FeCO$_3$)&    	0.00&	3.9&      116&	$\leq 0.00$ & 890&	1.04\\

PAH  (C$_{10}$H$_{14}$)& 	0.00&	1.0&    (178)&	$\leq 0.011$&	 N/A	&	1.04\\
\enddata
\tablenotetext{a}{Best-fit model $\chi^2_\nu=1.04$ with power law particle size distribution $dn/da \propto a^{-3.7}$,  5--35 $\mu$m range of fit, 336 degrees of freedom}
\tablenotetext{b}{Weight of the emissivity spectrum of each dust species required to match the G29-38 emissivity spectrum.}
\tablenotetext{c}{$N_{moles}(i)$ is the Density/Molecular Weight $\times$ Normalized Surface Area for mineral $i$. Errors are $\pm 15$\% (1$\sigma$).}
\tablenotetext{d}{All temperatures are $\pm 20$K (1$\sigma$)}
\tablenotetext{e}{We use the name niningerite to refer to Mg$_{x}$Fe$_{1-x}$S.
a niningerite composition of Mg$_{25}$Fe$_{75}$S may fit the data better.}
\tablenotetext{f}{Na$_{0.33}$Fe$_2$(Si,Al)$_4$O$_{10}$(OH)$_2\cdot3$H$_2$O)}
\end{deluxetable}

\clearpage

\begin{table}
\caption[]{Best-fitting moderately-optically-thin model}\label{modthicktab} 
\begin{flushleft} 
\begin{tabular}{lll}
Parameter & Best value & Confidence interval\\
\hline
$T_{vap}$ & 1100 K & 1050 -- 1200 K\\
$\tau_{\parallel}$ & 2 & 1 -- 8\\
$R_{min}$ & 50 & $\le 150$\\
$\alpha$ & 2.7 & 2.4 -- 2.9\\
\end{tabular}
\end{flushleft} 
\end{table}  

\clearpage

\def\extra{
\begin{figure}[th]
\plotone{f15}
\caption[silcompare]{
The relative abundance of olivine versus pyroxene determined from the infrared spectrum of G29-38, circumstellar clouds around 
HD 69830, HD 113766, HD 100546, and HD 163296, and comets 29P/Schwassmann-Wachmann 1 and 73P/Schwassmann-Wachmann 3.
The objects in the lower-left are young stars. 
G29-38 is at the upper-right, together with systems for which dust is inferred to be of asteroidal origin.
\label{silcompare}}
\end{figure}
}

\end{document}